\documentclass[fleqn,10pt]{article}
\usepackage{latexsym, graphicx, epsfig, amsmath, amssymb,amsfonts}
\usepackage{natbib,amsthm,version}
\usepackage{amsbsy,bm,multirow,enumerate}
\usepackage[titletoc,page]{appendix}
\usepackage[mathscr]{eucal}
\usepackage{mathtools}
\usepackage{color}
\usepackage{subfigure}
\usepackage[utf8]{inputenc}
\usepackage[english]{babel}
\usepackage{float}
\usepackage{caption}
\usepackage{systeme}
\usepackage[affil-it]{authblk}

\begin{document}

\title{ 
	Consistent and bound-preserving finite-volume WENO scheme for compressible two-/$N$-phase flows with Phase-Field mechanism
} 

\author{
	Ziyang Huang%
	\thanks{Email: \texttt{ziyanghuang@scut.edu.cn}; Corresponding author at the School of Marine Science and Engineering, Guangzhou International Campus, South China University of Technology, Guangzhou, Guangdong, China, 511400.}}

\affil{
	School of Marine Science and Engineering, Guangzhou International Campus, South China University of Technology, Guangzhou, Guangdong, China, 511400}


\maketitle


\begin{abstract}
In the present study, we propose a consistent and bound-preserving finite-volume WENO scheme that satisfies the requirements of consistency, conservation, equilibrium, and bound preservation for compressible multiphase flows with the Phase-Field mechanism.
The proposed WENO scheme is developed based on a new calculation of WENO weights determined by relative smoothness between stencils and on a coupled reconstruction of the masses and volume fractions.
Consistency of reduction and volume fraction summation to unity are considered during the development so that fictitious phases, local voids, or overfilling are not produced numerically when there are $N$ ($N \geqslant 1$) different immiscible phases.
The proposed WENO scheme is applied to the consistent and conservative Phase-Field method with adaptive mesh refinement enabled.
Various benchmark compressible two- and $N$-phase flows are performed to verify the properties of the proposed WENO scheme as well as its variant with the consistent limiter.
We finally demonstrate the capability of the proposed WENO scheme in shock-induced cavity collapse and shock–vessel–bubble interaction problems, with discussion of the necessity of bound preservation for high-order schemes and comparison of different compressible multiphase flow models.
\end{abstract}

\vspace{0.05cm}

\section{Introduction}\label{Sec:Introduction}
Weighted essentially non-oscillatory (WENO) schemes \citep{Liuetal1994,JiangShu1996} are a popular approach for problems with discontinuities or steep gradients, including high-speed compressible flows with shocks \citep{TitarevToro2004}, due to the ability of the WENO schemes to adaptively emphasize contributions from smooth stencils for discontinuity capturing, while recovering high order of accuracy in smooth regions. Although there are multiple variants of WENO schemes aiming to improve the performance, such as WENO-M \citep{Henricketal2005}, WENO-MDCD \citep{Martinetal2006}, WENO-Z \citep{Borgesetal2008}, and TENO \citep{Fuetal2016}, just to name a few, the WENO-JS scheme \citep{JiangShu1996} is still most widely used and serves as the building blog for these variants. Recent studies also consider bound preservation for a scalar and compressible single-phase flows when WENO schemes are implemented \citep{ZhangShu2010,ZhangShu2012}. Comprehensive reviews of WENO schemes are available in \citep{Shu1997,Shu1998,Shu2003,Shu2016,Shu2020}, and comparisons of some of the variants are available in \citep{MotheauWakefield2020}. 

\begin{figure}[!t]
	\centering
	\includegraphics[scale=.33]{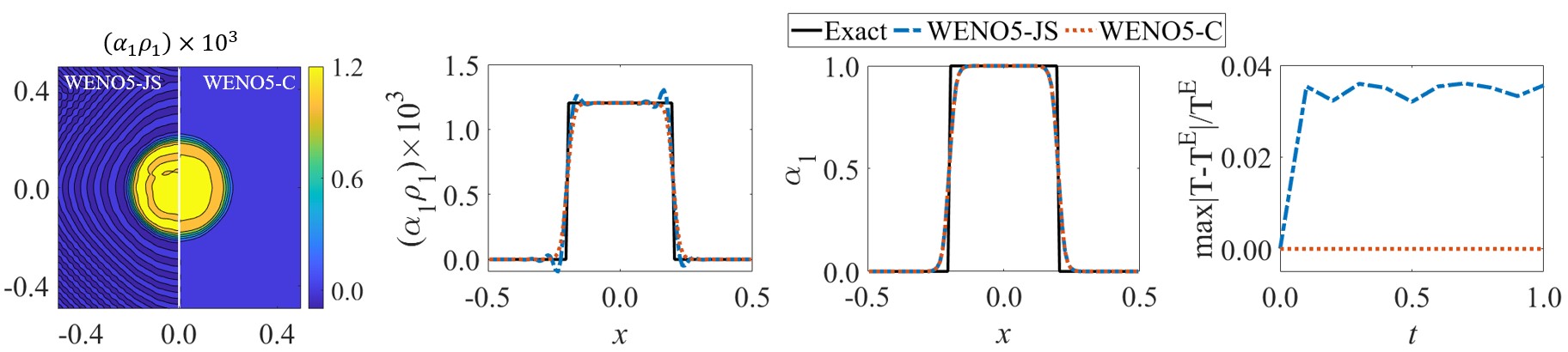}
	\caption{Contours of $(\alpha_1\rho_1)\times10^{3}$ after one period of advection (first panel), profiles of $(\alpha_1\rho_1)\times 10^3$ (second panel) and $\alpha_1$ (third panel) after one period of advection, and time history of temperature error, in the two-phase advection problem of \citep{HuangJohnsen2022} with the WENO5-JS \citep{JiangShu1996} and the proposed WENO5-C schemes.}\label{Fig:Advection}
\end{figure}

Given the wide range of applications in scientific and engineering problems, numerical models and approaches for compressible multiphase flows are actively studied. By contrast to methods that explicitly locate material interfaces, such as the front-tracking \cite{UnverdiTryggvason1992,Tryggvasonetal2001}, level-set \cite{OsherSethian1988,Sussmanetal1994,SethianSmereka2003,Gibouetal2018}, and volume-of-fluid (VOF) \cite{HirtNichols1981,ScardovelliZaleski1999,OwkesDesjardins2017} methods, the diffuse-interface capturing method \citep{SaurelPantano2018} has been actively developed to model compressible multiphase flows \citep{Abgrall1996,SaurelAbgrall1999,Allaireetal2002,Kapilaetal2001,Massonietal2002,PerigaudSaurel2005,CoralicColonius2014,JohnsenColonius2006,JohnsenHam2012,MovahedJohnsen2013,BeigJohnsen2015,HenrydeFrahanetal2015,Saureletal2008,Saureletal2009,Schmidmayeretal2017,FriessKokh2014,Petitpasetal2009,FriessKokh2014}, where material interfaces are treated as discontinuities that can be captured by schemes like WENO.
Under this framework, the implementation of WENO schemes in compressible multiphase flows first focused on satisfying the equilibrium requirement at isolated interfaces \citep{JohnsenColonius2006,JohnsenHam2012,CoralicColonius2014}. Recent progress further considered bound preservation \citep{ZhangCheng2022}, following the theoretical outcome for single-phase flows \citep{ZhangShu2010,ZhangShu2010Euler}, or implemented the WENO schemes as a limiter for discontinuous Galerkin schemes \citep{ZhangCheng2023,Whiteetal2025}.
In compressible multiphase flows, the appearance of material interfaces and their interactions with non-linear waves add extra complexity, and a direct implementation of current WENO schemes can introduce unexpected errors.
Fig.~\ref{Fig:Advection} shows errors generated by the WENO-JS scheme \citep{JiangShu1996} in its direct application to the two-phase advection problem in \citep{HuangJohnsen2022} (detailed in Section~\ref{Sec:Advection-TwoPhase}), where an air bubble (Phase~1) is translated in water (Phase~2) after one period of advection. It is expected that the WENO-JS scheme is able to detect discontinuities and produce non-oscillatory results, which is true for the volume fraction ($\alpha_1$) but not true for the corresponding mass ($\alpha_1\rho_1$), although the velocity and pressure maintain their uniformity. The WENO-JS scheme does not maintain thermal equilibrium either, resulting in a significant temperature error.
Moreover, the existing implementation of WENO schemes is limited to two-phase flows. When there are more than two phases, consistency of reduction \citep{BoyerMinjeaud2014,Dong2018,Huangetal2020N,Huangetal2020B} and volume fraction (or mass fraction) summation to unity are not directly satisfied by the WENO schemes \citep{Huangetal2020N,BaumgartBlanquart2024}, resulting in the production of fictitious phases, local voids, or overfilling. 
In addition, material interfaces, as linearly degenerate waves, are inevitably thickened over time by numerical diffusion implied in the capturing schemes \citep{Harten1977,Harten1978,SaurelPantano2018}, resulting in numerical mixing between different materials/phases. Recent studies suggest introducing the Phase-Field mechanism to counteract numerical diffusion \citep{Shuklaetal2010,Tiwarietal2013,Jainetal2020,HuangJohnsen2022,HuangJohnsen2023}. However, incorporation of WENO schemes into this additional Phase-Field mechanism has yet been performed.
Therefore, WENO schemes must be tailored to compressible multiphase flows, so that relevant physics can be accurately simulated.

In the present study, the behavior of the WENO-JS scheme \citep{JiangShu1996} in compressible multiphase flows is first analyzed to understand its errors shown in Fig.~\ref{Fig:Advection}, which results in a new calculation of the non-linear WENO weights based on \textit{relative} smoothness between stencil candidates. Following the consistency, conservation, equilibrium, and bound preservation requirements analyzed in \citep{HuangJohnsen2024} for finite-volume reconstructions in compressible multiphase flows, we apply the modified WENO weights and develop a consistent and bound-preserving WENO reconstruction scheme. To the best of our knowledge, this is the first WENO scheme that satisfies the four critical requirements for compressible multiphase flows, and thus the unphysical behaviors of the WENO-JS scheme are not observed, see the WENO-C results in Fig.~\ref{Fig:Advection}.
The proposed WENO scheme is developed under a general setup that there are $N$ ($N \geqslant 1$) different immiscible phases, and both consistency of reduction and volume fraction summation to unity are taken into account during the development.
An alternative approach with the consistent limiter \citep{HuangJohnsen2023} to achieve these properties is also discussed.
The present analysis and modifications are not limited to WENO-JS and can readily be adapted to its variants like \citep{Henricketal2005,Martinetal2006,Borgesetal2008,Fuetal2016}.
The proposed WENO scheme is implemented in the consistent and conservative Phase-Field method with adaptive mesh refinement \citep{HuangJohnsen2022,HuangJohnsen2023,HuangJohnsen2024,Huangetal2025} for compressible multiphase flows, and various two- and $N$-phase flows are investigated and discussed to demonstrate the proposed approach.

The remainder of this paper is organized as follows.
In Section~\ref{Sec:Preliminary}, the governing equations and the numerical framework are described, along with a summary of the requirements for a finite-volume reconstruction in compressible multiphase flows.
In Section~\ref{Sec:WENO}, the WENO-JS scheme \citep{JiangShu1996} is analyzed, resulting in a modification to the WENO weights and the development of the proposed consistent and bound-preserving WENO scheme (``WENO-C'') and its variant (``WENO-CL'').
In Section~\ref{Sec:Results}, numerical tests and applications are performed to demonstrate the proposed WENO scheme with discussions.
Finally, the present study concludes in Section~\ref{Sec:Conclusions}.

\section{Consistent and conservative Phase-Field method for compressible multiphase flows}\label{Sec:Preliminary}
\subsection{Governing equations}\label{Sec:Eulerian}
With the Phase-Field mechanism for the phases, denoted by $\{\mathbf{J}_p\}_{p=1}^N$, and satisfying $\sum_{p=1}^N \mathbf{J}_p=\mathbf{0}$ (or $\sum_{p=1}^N \nabla \cdot \mathbf{J}_p=0$), we obtain the $N$-phase $(N \geqslant 1)$ Euler/Phase-Field model simplified from the complete model derived in \citep{HuangJohnsen2022}, which reads
\begin{equation}\label{Eq:Eulerian}
\begin{split}
&\frac{\partial \mathbf{U}}{\partial t}
+
\nabla \cdot \mathbf{F}^{HB}
=
\mathbf{H}^{HB} \nabla \cdot \mathbf{u}
+
\nabla \cdot \mathbf{F}^{PF}\left[\{\mathbf{J}_p\}_{p=1}^N\right],\\
&\mathbf{U}
=
\left[ \{(\alpha_p \rho_p)\}_{p=1}^N,(\rho \mathbf{u}),(\rho E),\{\alpha_p\}_{p=1}^N\right]^T,\\
&\mathbf{F}^{HB}
=
\left[\{\mathbf{u} (\alpha_p \rho_p)\}_{p=1}^N,(\rho \mathbf{u}) \mathbf{u} + P \mathrm{I},\mathbf{u} ( (\rho E)+P),\{\mathbf{u} \alpha_p\}_{p=1}^N\right]^T,\\
&\mathbf{H}^{HB}
=
\left[\{0\}_{p=1}^N,\mathbf{0},0,\{ \alpha_p (1+K_p) \}_{p=1}^N\right]^T,\\
&\mathbf{F}^{PF}\left[\{\mathbf{J}_p\}_{p=1}^N\right]
=
\left[\{\mathbf{J}_p \rho_p\}_{p=1}^N, \sum_{p=1}^N \mathbf{J}_p \rho_p \mathbf{u}, \sum_{p=1}^N \mathbf{J}_p\rho_p \left(\frac{\mathbf{u} \cdot \mathbf{u}}{2} + e_p\right), \{\mathbf{J}_p\}_{p=1}^N\right]^T,
\end{split}
\end{equation}
where $\mathbf{U}$ is the vector of conservative variables consisting of the phasic masses $\{(\alpha_p\rho_p)\}_{p=1}^N$, momentum $(\rho \mathbf{u})$, total energy $(\rho E)$, and phasic volume fractions $\{\alpha_p\}_{p=1}^N$;
$\mathbf{F}^{HB}$, $\mathbf{H}^{HB} \nabla \cdot \mathbf{u}$, and $\mathbf{F}^{PF}$ are the hyperbolic flux vector, the hyperbolic source vector, and the Phase-Field flux vector, respectively. Viscous stresses and heat conduction are not considered.
Eq.~(\ref{Eq:Eulerian}) satisfies mass, momentum, and energy conservation, the second law of thermodynamics, the Galilean invariance (which implies the kinematic, mechanical, and thermal equilibrium at isolated interfaces), and consistency of reduction, as analyzed in \citep{HuangJohnsen2022}.

The phasic quantities are related to the mixture quantities via the following mixture rules: $\rho=\sum_{p=1}^N (\alpha_p\rho_p)$ and $(\rho e)=(\rho E)-\frac{1}{2} \rho \mathbf{u} \cdot \mathbf{u}=\sum_{p=1}^N \alpha_p \rho_p e_p$, where $\rho$ is the mixture density, $\mathbf{u}$ is the flow velocity, and $e_p$ is the specific internal energy of phase~$p$.
To close the system, the specific internal energy of each phase is related to the thermodynamic pressure $P$ via a pressure-based equation of state, i.e., $e_p=\mathsf{e}_p(P,\rho_p)$, and related to the phasic temperature $T_p$ via a temperature-based equation of state, i.e., $e_p=\mathsf{e}_p^T(T_p,\rho_p)$. The present five-equations compressible multiphase model \citep{HuangJohnsen2022} assumes equilibrium pressure between the phases in mixture regions.
The phasic and mixture sound speeds are obtained from
\begin{equation}\label{Eq:SoundSpeed}
\rho_p c_p^2 = \frac{\frac{P}{\rho_p}-\rho_p \left(\frac{\partial e_p}{\partial \rho_p}\right)_P}{\left(\frac{\partial e_p}{\partial P}\right)_{\rho_p}},
\quad
\rho c^2 = \frac{\sum_{p=1}^N \left(\frac{\partial e_p}{\partial P}\right)_{\rho_p} (\alpha_p\rho_p) (\rho_p c_p^2) (1+K_p)}{\sum_{p=1}^N \left(\frac{\partial e_p}{\partial P}\right)_{\rho_p} (\alpha_p\rho_p)},
\end{equation}
respectively, where $\{K_p\}_{p=1}^N$ need to satisfy $\sum_{p=1}^N \alpha_p K_p=0$ to preserve $\sum_{p=1}^N \alpha_p =1$ on the model level. The present study considers both $\{K_p=0\}_{p=0}^N$ (the model of Allaire et al. \citep{Allaireetal2002}) and $\{K_p=\frac{\rho c^2}{\rho_p c_p^2}-1\}_{p=1}^N$ (the model of Kapila et al. \citep{Kapilaetal2001}).

\subsection{Numerical approach}\label{Sec:Scheme}
\subsubsection{General framework}\label{Sec:Scheme-Overview}
To solve Eq.~(\ref{Eq:Eulerian}) numerically with different formulations of the Phase-Field mechanism, we follow our previous approach \citep{HuangJohnsen2022,HuangJohnsen2023,HuangJohnsen2024,Huangetal2025}, based on a finite volume discretization and fractional time stepping.
In each time step, we first perform the hyperbolic step that advances
\begin{equation}\label{Eq:Hyperbolic}
\frac{\partial \overline{\mathbf{U}}}{\partial t}
+
\nabla \cdot \hat{\mathbf{F}}^{HB}\left(\mathbf{U}^L,\mathbf{U}^R\right)
=
\mathbf{H}^{HB}\left(\overline{\mathbf{U}}\right) \nabla \cdot \hat{\mathbf{u}},
\end{equation}
where $\overline{\mathbf{U}}$ denotes the cell-averaged data, $\mathbf{U}^{L,R}$ is the reconstructed data at two sides of a cell face, $\hat{\mathbf{F}}^{HB}$ represents an approximate Riemann solver, and $\hat{\mathbf{u}}$ is an approximation of velocity at cell faces. It is critical to relate $\hat{\mathbf{u}}$ to $\hat{\mathbf{F}}^{HB}$ via $\hat{\mathbf{u}}=\hat{\mathbf{F}}^{HB,\alpha_p}\left(\mathbf{U}^L|_{\alpha_p^L = 1},\mathbf{U}^R|_{\alpha_p^R = 1}\right)$, where $\hat{\mathbf{F}}^{HB,\xi}$ denotes the numerical flux of $\xi$, to preserve the consistency of reduction (no fictitious phase) and $\sum_{p=1}^N \alpha_p=1$ (no local void or overfilling) \citep{HuangJohnsen2022,HuangJohnsen2023}. The outcome of the hyperbolic step is $\overline{\mathbf{U}}^{HB}$.

Then, using the volume fractions from $\overline{\mathbf{U}}^{HB}$ as the order parameters $\{\phi_p\}_{p=1}^N$, the selected Phase-Field mechanism is calculated, which later is mapped to auxiliary variables $\{Q_p\}_{p=1}^N$ via
\begin{equation}\label{Eq:Q-E}
\nabla \cdot \left( 
\phi_p \left(1 - \phi_p\right) \nabla Q_p 
\right)
=
\widehat{[\nabla \cdot {\mathbf{J}}_p]},
\quad
1 \leqslant p \leqslant N,
\end{equation}
where $\widehat{[\nabla \cdot {\mathbf{J}}_p]}$ is a numerical approximation of $[\nabla \cdot {\mathbf{J}}_p]$. From Eq.~(\ref{Eq:Q-E}), the effect of the Phase-Field mechanism is extracted by $\{\nabla Q_p\}_{p=1}^N$.

To end the time step, the Phase-Field step is performed, which advances
\begin{equation}\label{Eq:PhaseField}
\frac{\partial \overline{\mathbf{U}}}{\partial t}
=
\nabla \cdot \hat{\mathbf{F}}^{PF}\left(\mathbf{U}^L,\mathbf{U}^R;\nabla Q\right),
\end{equation}
where $\hat{\mathbf{F}}^{PF}$ is the numerical Phase-Field flux vector.
The current framework is flexible and can accommodate different formulations of the Phase-Field mechanism by appropriately modifying the right-hand side of Eq.~(\ref{Eq:Q-E}).

\subsubsection{Requirements for reconstructed data}\label{Sec:Scheme-Requirement}
To achieve consistency, equilibrium, conservation, and admissibility (bound preservation) with the scheme in Section~\ref{Sec:Scheme-Overview}, the following important requirements for the data reconstruction needs to be satisfied, which are critical for the success of simulating compressible multiphase flows.

\textit{Consistency between mass and volume fraction} \citep{HuangJohnsen2022,HuangJohnsen2023,HuangJohnsen2024} requires $(\alpha_p\rho_p)^{L,R}=\rho_p \times \alpha_p^{L,R}$ when $\rho_p$ is constant in the reconstruction stencil (for any $p$). In compressible multiphase flows, material interfaces are represented by both the transition of volume fraction ($\alpha_p$) from $0$ to $1$ and the transition of mass ($\alpha_p\rho_p$) from $0$ to a positive number. This consistency requirement ensures that the volume fraction transition and the mass transition coincide. If not, a singular phasic density ($\rho_p$) may be produced, manifesting as unphysical mass spikes near interfaces \citep{HuangJohnsen2022}, and may further cause simulation failures in large-density-ratio problems.

\textit{Kinematic, mechanical, and thermal equilibria at isolated interfaces} imply that the normal velocity is continuous across isolated interfaces, that there is no pressure jump in the absence of any surface forces like surface tension, and additional that the temperature is continuous if neighboring phases have the same temperature. To satisfy this requirement, a well-accepted approach \citep{Abgrall1996,SaurelAbgrall1999,JohnsenColonius2006,CoralicColonius2014,BeigJohnsen2015,HuangJohnsen2022,HuangJohnsen2023} is to reconstruct the primitive variables $\mathbf{V}=\left[\{(\alpha_p\rho_p)\}_{p=1}^N,\mathbf{u},P,\{\alpha_p\}_{p=1}^N\right]^T$, resulting in $\mathbf{U}^{L,R}=\psi_{\mathbf{V} \rightarrow \mathbf{U}}\left(\mathbf{V}^{L,R}\right)$, where $\mathbf{V}^{L,R}$ are the reconstructed primitive variables and $\psi_{\mathbf{V}\rightarrow\mathbf{U}}(\cdot)$ denotes the mapping of the primitive variables to the conservative variables. Furthermore, recent analysis in \citep{HuangJohnsen2024} shows that the consistency requirement mentioned above is also needed to satisfy this equilibrium requirement for general equations of state.

The reconstructed functions of the conservative variables resulting from the reconstructed primitive variables still need to satisfy the \textit{conservation} requirement \citep{HenrydeFrahanetal2015,HuangJohnsen2024}, i.e., the cell-averaged values of these reconstructed functions must match the given cell-averaged conservative variables in each grid cell. For piecewise polynomial reconstructions of order $N_{poly}$ performed dimension-by-dimension, we can relate the given cell-averaged data to the reconstructed polynomials with the Gauss-Lobatto quadrature rule of order $N_{quad}$ ($\geqslant (N_{poly}+3)/2$) in each grid cell:
\begin{equation}\label{Eq:CellAverage}
\overline{\mathbf{U}}_i
=
\sum_{n_{quad}=1}^{N_{quad}} \hat{\omega}^{(n_{quad})} \mathbf{U}_i\left(\hat{x}_i^{(n_{quad})}\right)
=
\hat{\omega}^{(1)} \mathbf{U}_{i-1/2}^R
+
\hat{\omega}^{(N_{quad})} \mathbf{U}_{i+1/2}^L
+
\left(1-\hat{\omega}^{(1)}-\hat{\omega}^{(N_{quad})}\right) \mathbf{U}_i^C,
\end{equation}
where $\overline{\mathbf{U}}_i$ is the cell-averaged data given at cell $I_i=[x_{i-1/2},x_{i+1/2}]$, $\mathbf{U}_i(x)$ is the reconstructed polynomials of the conservative variables at the cell, and $\left\{\hat{x}_i^{(n_{quad})},\hat{\omega}^{(n_{quad})}\right\}_{n_{quad}=1}^{N_{quad}}$ are the quadrature points in $[x_{i-1/2},x_{i+1/2}]$ and weights in $[-1/2,1/2]$.
The reconstructed values at the cell faces are $\mathbf{U}_{i-1/2}^R=\mathbf{U}_{i}\left(\hat{x}_i^{(1)}\right)=\mathbf{U}_{i}(x_{i-1/2})$ and $\mathbf{U}_{i+1/2}^L=\mathbf{U}_{i}\left(\hat{x}_i^{(N_{quad})}\right)=\mathbf{U}_{i}(x_{i+1/2})$, and the values at other quadrature points inside the cell are combined to be $\mathbf{U}_i^C = \sum_{n_{quad}=2}^{N_{quad}-1} \frac{\hat{\omega}^{(n_{quad})} \mathbf{U}_i\left(\hat{x}_i^{(n_{quad})}\right)}{1-\hat{\omega}^{(1)}-\hat{\omega}^{(N_{quad})}}$.

Lastly, it is critical that the solution of the scheme in Section~\ref{Sec:Scheme-Overview} remains \textit{admissible} or \textit{bound-preserving} \citep{HuangJohnsen2024}. For equations of state in the form
\begin{equation}\label{Eq:EOS-Bound-General}
\mathsf{e}_p(P,\rho_p)=\frac{\mathsf{e}_p^{(1)}(P)+\rho_p \mathsf{e}_p^{(2)}(P)}{\rho_p},
\end{equation}
where $\mathsf{e}_p^{(1)}(P)$ and $\mathsf{e}_p^{(2)}(P)$ are monotonically increasing functions of pressure only, if the reconstructed data $\mathbf{U}^{L,R,C} \in \mathcal{U}$, then the solution of the scheme in Section~\ref{Sec:Scheme-Overview} is still in $\mathcal{U}$ with the bound-preserving numerical flux vectors and suitable CFL conditions for the hyperbolic and Phase-Field steps \citep{HuangJohnsen2024}. Here, $\mathcal{U}$ is the admissible set defined as
\begin{equation}\label{Eq:AdmissibleSet}
\mathcal{U}
=
\left\{
\mathbf{U} \in \mathbb{R}^{2N+d+1}
\left|
\{(\alpha_p\rho_p)\}_{p=1}^N \in [0,+\infty),
\{\alpha_p\}_{p=1}^N \in [0,1],
\sum_{p=1}^N \alpha_p=1,
\psi(\mathbf{U}) \in [0,+\infty)
\right.
\right\},
\quad
d=1,2,3,
\end{equation}
which includes (1) mass positivity, (2) volume fraction boundedness, (3) volume fraction summation to unity, and (4) the energy constraint
\begin{equation}\label{Eq:EnergyBound}
\psi(\mathbf{U})
=
(\rho E) - \frac{1}{2} \frac{(\rho \mathbf{u})\cdot(\rho \mathbf{u})}{\sum_{p=1}^N(\alpha_p\rho_p)}
-
\sum_{p=1}^N (\alpha_p\rho_p) \mathsf{e}_p\left( (P_{\min})_p,\rho_p\right),
\end{equation}
where $\{(P_{\min})_p\}_{p=1}^N$ are user-prescribed phasic minimum pressure. The admissible set $\mathcal{U}$ in Eq.~(\ref{Eq:AdmissibleSet}) is convex with equations of state in the category of Eq.~(\ref{Eq:EOS-Bound-General}). As shown in \citep{HuangJohnsen2024}, the conservation requirement mentioned above is also needed to satisfy this admissibility (bound preservation), reflected in the requirement that not only the reconstructed data at cell faces ($\mathbf{U}^{L,R}$) but also the reconstructed data inside grid cells ($\mathbf{U}^C$) belong to $\mathcal{U}$.
We further note that the consistency requirement can be relaxed for the kinematic and mechanical equilibria if equations of state have the form of Eq.~(\ref{Eq:EOS-Bound-General}), while it is still needed for thermal equilibrium \citep{HuangJohnsen2024}.

\section{WENO reconstruction scheme}\label{Sec:WENO}
In this section, we first provide the general formulation of the WENO reconstruction scheme, followed by an analysis of the WENO-JS scheme \citep{JiangShu1996} to understand its errors in compressible multiphase flows, as shown in Fig.~\ref{Fig:Advection}. Then, a modification to the WENO weights is proposed. Finally, the consistent and bound-preserving WENO scheme that satisfies all the requirements in Section~\ref{Sec:Scheme-Requirement} for compressible multiphase flows is developed and analyzed, followed by discussions of its variant with the consistent limiter \citep{HuangJohnsen2024} and its implementation on adaptive meshes.

\subsection{Overview of WENO}\label{WENO-Overview}
Following \citep{JiangShu1996,Shu1998,Shu2020}, given the cell-averaged values of a scalar function $f(x)$ in stencil $S=\{i-r,i-r+1,...,i+r-1,i+r\}$, the $(2r+1)$th-order WENO scheme reconstructs the values of $f$ at $x_{i+1/2}$ and $x_{i-1/2}$ in cell $I_i=[x_{i-1/2},x_{i+1/2}]$, denoted by $f_{i+1/2}^L$ and $f_{i-1/2}^R$, respectively. Due to symmetry, we discuss the reconstruction of $f_{i+1/2}^L$; a similar process can be followed for $f_{i+1/2}^R$.

Using stencil $S$, a $(2r+1)$th-order polynomial is reconstructed, whose value at $x_{i+1/2}$ is $f_{i+1/2}^{(S)}$. Moreover, there are $N_s$ $(=r+1)$ smaller stencils within $S$ that include cell $I_i$, i.e., $s=\{i-r+s-1,i-r+s,...,i+s-2,i+s-1\}$ for $s=1$ to $s=N_s$. For stencil $s$, a $(r+1)$th-order polynomial is reconstructed, whose value at $x_{i+1/2}$ is $f_{i+1/2}^{(s)}$. As a result, a set of ideal weights $\{d_{i+1/2}^{(s)}\}_{s=1}^{N_s}$ is obtained by satisfying $f_{i+1/2}^{(S)}=\sum_{s=1}^{N_s} d_{i+1/2}^{(s)} \times f_{i+1/2}^{(s)}$. Furthermore, the smoothness of the $(r+1)$th-order polynomial of $f$ reconstructed from stencil $s$ is measured by $IS_{i}^{(s)}[f]$. For simplicity, we call $IS_{i}^{(s)}[f]$ the smoothness indicator of stencil $s$. Finally, the reconstructed $f_{i+1/2}^L$ is
\begin{equation}\label{Eq:WENO}
f_{i+1/2}^L=\sum_{s=1}^{N_s} \omega_{i+1/2}^{(s)}[f] \times f_{i+1/2}^{(s)},
\end{equation}
where $\omega_{i+1/2}^{(s)}[f]$ is the WENO weight for stencil $s$ using $d_{i+1/2}^{(s)}$ and $IS_{i}^{(s)}[f]$ such that non-smooth stencils are excluded.

\subsection{Analysis of WENO-JS}\label{Sec:WENO-JS}
We first analyze the WENO-JS scheme \citep{JiangShu1996} to understand the errors shown in Fig.~\ref{Fig:Advection}. For a clear presentation, we focus on the third-order ($r=1$) case, and the conclusion applies to the higher-order WENO-JS scheme.

Denoting $\Delta x_i=x_{i+1/2}-x_{i-1/2}$, the third-order WENO-JS scheme has
\begin{equation}\label{Eq:WENO-JS-Stencil}
\begin{split}
f_{i+1/2}^{(1)}=\overline{f}_i+\frac{\Delta x_i}{\Delta x_{i} + \Delta x_{i-1}}(\overline{f}_{i}-\overline{f}_{i-1}),&
\quad
f_{i+1/2}^{(2)}=\overline{f}_i+\frac{\Delta x_i}{\Delta x_{i+1}+\Delta x_i}(\overline{f}_{i+1}-\overline{f}_{i}),\\
d_{i+1/2}^{(1)}=\frac{\Delta x_{i+1}}{\Delta x_{i+1} + \Delta x_i + \Delta x_{i-1}},&
\quad
d_{i+1/2}^{(2)}=\frac{\Delta x_i + \Delta x_{i-1}}{\Delta x_{i+1} + \Delta x_i + \Delta x_{i-1}},\\
{IS}_{i}^{(1)}[f]=\frac{4(\Delta x_i)^2}{(\Delta x_i + \Delta x_{i-1})^2}(\overline{f}_i-\overline{f}_{i-1})^2,&
\quad
{IS}_{i}^{(2)}[f]=\frac{4 (\Delta x_i)^2}{(\Delta x_{i+1} + \Delta x_i)^2}(\overline{f}_{i+1}-\overline{f}_i)^2,
\end{split}
\end{equation}
and the corresponding WENO weights are
\begin{equation}\label{Eq:WENO-JS-Weight}
\omega_{i+1/2}^{(1)}[f](\varepsilon)=\frac{\frac{d_{i+1/2}^{(1)}}{(IS_i^{(1)}[f]+\varepsilon)^2}}{\frac{d_{i+1/2}^{(1)}}{(IS_i^{(1)}[f]+\varepsilon)^2}+\frac{d_{i+1/2}^{(2)}}{(IS_i^{(2)}[f]+\varepsilon)^2}},
\quad
\omega_{i+1/2}^{(2)}[f](\varepsilon)=\frac{\frac{d_{i+1/2}^{(2)}}{(IS_i^{(2)}[f]+\varepsilon)^2}}{\frac{d_{i+1/2}^{(1)}}{(IS_i^{(1)}[f]+\varepsilon)^2}+\frac{d_{i+1/2}^{(2)}}{(IS_i^{(2)}[f]+\varepsilon)^2}},
\end{equation}
where $\varepsilon$ is a small number to avoid division by zero. As a result, the reconstructed value of $f$ becomes
\begin{equation}\label{Eq:WENO-JS}
f_{i+1/2}^L(\varepsilon)=\omega_{i+1/2}^{(1)}[f](\varepsilon) \times f_{i+1/2}^{(1)} + \omega_{i+1/2}^{(2)}[f](\varepsilon) \times f_{i+1/2}^{(2)},
\end{equation}
depending on $\varepsilon$ as well.

Considering another scalar function $g(x)=af(x)+b$, where $a$ and $b$ are constant, the WENO-JS scheme has $g_{i+1/2}^{(s)}=af_{i+1/2}^{(s)}+b$, while ${IS}_i^{(s)}[g]=a^2 {IS}_i^{(s)}[f]$ (see Eq.~(\ref{Eq:WENO-JS-Stencil})), resulting in $\omega_{i+1/2}^{(s)}[g](\varepsilon)=\omega_{i+1/2}^{(s)}[f](\varepsilon/a^2)$ (see Eq.~(\ref{Eq:WENO-JS-Weight})). Therefore, the reconstructed value of $g$ is related to that of $f$ in Eq.~(\ref{Eq:WENO-JS}) by
\begin{equation}\label{Eq:WENO-JS-Linear}
g_{i+1/2}^L(\varepsilon)=a f_{i+1/2}^L\left(\frac{\varepsilon}{a^2}\right) + b.
\end{equation}
We note that Eq.~(\ref{Eq:WENO-JS-Linear}) is generally true for the WENO-JS scheme independent of the order of accuracy.

To explain the errors of the WENO-JS scheme shown in Fig.~\ref{Fig:Advection}, we consider $f=\alpha_1$ and $g=(\alpha_1\rho_1)$, resulting in $a=\rho_1=1.204\times10^{-3}$ and $b=0$.
On one hand, it is learned from Eq.~(\ref{Eq:WENO-JS-Linear}) that the WENO-JS scheme fails the consistency requirement in Section~\ref{Sec:Scheme-Requirement}, i.e., $(\alpha_1\rho_1)_{i+1/2}^L \neq \rho_1 \times (\alpha_1)_{i+1/2}^L$, which manifests in the temperature error in Fig.~\ref{Fig:Advection}. Such an inconsistency can also introduce velocity and pressure errors at isolated interfaces with general equations of state as analyzed and verified in \citep{HuangJohnsen2024} and produce unphysical mass spikes \citep{HuangJohnsen2022} that lead to failure in large-density-ratio problems.
On the other hand, the WENO-JS scheme may lose its essentially non-oscillatory behavior when $a$ is small. Although $\varepsilon$ is sufficiently small compared to $IS_i^{(s)}[f]$ in Eq.~(\ref{Eq:WENO-JS-Weight}) such that it has no effect on deactivating the non-smooth stencils of $f$, $\varepsilon/a^2$ is not necessarily a small number, resulting in failures to exclude the non-smooth stencils of $g=af+b$. In the two-phase advection example shown in Fig.~\ref{Fig:Advection}, $\varepsilon/a^2=\varepsilon/\rho_1^2=10^{-6}/(1.204\times10^{-3})^2=0.6898$, and the non-smooth stencils of $(\alpha_1\rho_1)$ can still be included in the reconstruction. Therefore, $(\alpha_1\rho_1)$ from the WENO-JS scheme is oscillatory, although the corresponding $\alpha_1$ is essentially non-oscillatory, as shown in Fig.~\ref{Fig:Advection}.

\subsection{Modified WENO weights}\label{Sec:WENO-W}
In the present study, we propose a modified approach to compute the WENO weights so that the essentially non-oscillatory behavior is preserved for both $f(x)$ and $g(x)=af(x)+b$. Our idea is inspired by the fact that the stencil selection mechanism in the WENO scheme is relative: stencil $s$ has larger WENO weights and thus greater contributions to the reconstructed value than stencil $k$ if stencil $s$ is smoother. To quantify this relative smoothness between stencils $s$ and $k$ of cell $I_i$, we define $IS_{i}^{(s,k)}[f]$ for $f(x)$, and use it to calculate the proposed modified WENO weights
\begin{equation}\label{Eq:WENO-W-Weight}
\omega_{i+1/2}^{(s)}[f]
=
\frac{d_{i+1/2}^{(s)}}{
d_{i+1/2}^{(s)} +\sum_{k=1,k \neq s}^{N_s} d_{i+1/2}^{(k)} (IS_i^{(s,k)}[f])^2
},
\quad
IS_i^{(s,k)}[f]=\left\{\begin{array}{cc}
     1,&  IS_i^{(s)}[f]=IS_i^{(k)}[f]=0,\\
     IS_i^{(s)}[f]/IS_i^{(k)}[f],& \mathrm{else}.
\end{array}
\right.
\end{equation}
When $IS_i^{(s)}[f]$ and $IS_i^{(k)}[f]$ are both zero, stencils $s$ and $k$ are equally smooth, and therefore $IS_i^{(s,k)}[f]=1$. It is clear that Eq.~(\ref{Eq:WENO-W-Weight}) does not exclude any smooth stencils, and thus the proposed modified WENO weights do not affect the order of accuracy.
It should also be noted that there is no need to introduce $\varepsilon$ in the modified WENO weights.

Considering again $g(x)=af(x)+b$, since $IS_i^{(s)}[g]=a^2 IS_i^{(s)}[f]$ from Eq.~(\ref{Eq:WENO-JS-Stencil}), we now have $IS_i^{(s,k)}[g]=IS_i^{(s,k)}[f]$, resulting in $g_{i+1/2}^L=a f_{i+1/2}^L+b$ with the modified WENO weights in Eq.(\ref{Eq:WENO-W-Weight}). As a result, the non-smooth stencils of $g(x)=af(x)+b$ can be directly sensed by the modified WENO weights no matter how small $a$ is. This property is important for compressible multiphase flows, as the phasic density can become very small under a strong expansion.
Hereafter, the WENO scheme for which the weights are computed from Eq.~(\ref{Eq:WENO-W-Weight}) is named ``WENO-W''.

\subsection{Consistent and bound-preserving WENO scheme for compressible multiphase flows}\label{Sec:WENO-C}
Although the proposed modified WENO weights in Eq.~(\ref{Eq:WENO-W-Weight}) result in $g_{i+1/2}^L=af_{i+1/2}^L+b$ given $g(x)=af(x)+b$ to satisfy the consistency requirement, our practice indicates that this property is sensitive to and can be quickly deteriorated by round-off error. Additionally, the equilibrium, conservation, and admissibility (bound preservation) requirements listed in Section~\ref{Sec:Scheme-Requirement} for compressible multiphase flows are not directly satisfied by the WENO scheme.
To address these issues, we further modify the WENO scheme to preserve all the important requirements for compressible multiphase flows.

To satisfy the equilibrium requirement, we follow the approach \citep{JohnsenColonius2006,CoralicColonius2014,BeigJohnsen2015} that the WENO reconstruction is performed on the primitive variables $\overline{\mathbf{V}}=\psi_{\mathbf{U} \rightarrow \mathbf{V}}(\overline{\mathbf{U}})$, where $\psi_{\mathbf{U} \rightarrow \mathbf{V}}(\cdot)$ is the mapping of the conservative to primitive variables. As a result, $\mathbf{u}$ and $P$ are reconstructed individually with the WENO-W scheme, while the reconstructions of $\{(\alpha_p\rho_p)\}_{p=1}^N$ and $\{\alpha_p\}_{p=1}^N$ are coupled. In the following, we only present the formulations for $(\alpha_p\rho_p)_{i+1/2}^L$ and $(\alpha_p)_{i+1/2}^L$, and the same formulations are applied to $(\alpha_p\rho_p)_{i-1/2}^R$ and $(\alpha_p)_{i-1/2}^R$.

The first step is to obtain $(\alpha_p\rho_p)_{i+1/2}^L$ and $(\alpha_p)_{i+1/2}^L$ from
\begin{equation}\label{Eq:WENO-C-MassVolume}
(\alpha_p\rho_p)_{i+1/2}^L=\sum_{s=1}^{N_s} \omega_{i+1/2}^{(s)}[(\alpha_p\rho_p)] \times (\alpha_p\rho_p)_{i+1/2}^{(s)},
\quad
(\alpha_p)_{i+1/2}^L=\sum_{s=1}^{N_s} \omega_{i+1/2}^{(s)}[(\alpha_p\rho_p)] \times (\alpha_p)_{i+1/2}^{(s)},
\end{equation}
where $\omega_{i+1/2}^{(s)}[(\alpha_p\rho_p)]$ is the modified WENO weight in Eq.~(\ref{Eq:WENO-W-Weight}) based on $(\alpha_p\rho_p)$.
To enforce the consistency requirement \citep{HuangJohnsen2022,HuangJohnsen2023,HuangJohnsen2024}, $\omega_{i+1/2}^{(s)}[(\alpha_p\rho_p)]$ is used to calculate both $(\alpha_p\rho_p)_{i+1/2}^L$ and $(\alpha_p)_{i+1/2}^L$ in Eq.~(\ref{Eq:WENO-C-MassVolume}).
We choose $\omega_{i+1/2}^{(s)}[(\alpha_p\rho_p)]$ because it captures both discontinuities on $(\alpha_p\rho_p)$ appearing in both bulk-phase (e.g. due to shocks) and interfacial (material interfaces) regions, while $\omega_{i+1/2}^{(s)}[\alpha_p]$ can only sense material interfaces.
As $\alpha_p$ is either $0$ or $1$ in bulk-phase regions, its reconstructed value at these locations is not affected by the WENO weight values.
After the modification to the WENO weights in Section~\ref{Sec:WENO-W}, material interfaces that can be detected by $\omega_{i+1/2}^{(s)}[\alpha_p]$ can also be detected by $\omega_{i+1/2}^{(s)}[(\alpha_p\rho_p)]$. 

The second step is to enforce the positivity of $(\alpha_p\rho_p)$ and $\alpha_p$ via
\begin{equation}\label{Eq:WENO-C-Positivity}
(\alpha_p\rho_p)_{i+1/2}^L \leftarrow \overline{(\alpha_p\rho_p)}_i + (\theta_p)_{i} \left( (\alpha_p\rho_p)_{i+1/2}^L - \overline{(\alpha_p\rho_p)}_i \right),
\quad
(\alpha_p)_{i+1/2}^L \leftarrow \overline{(\alpha_p)}_i + (\theta_p)_{i} \left( (\alpha_p)_{i+1/2}^L-\overline{(\alpha_p)}_i \right),
\end{equation}
where $\theta_{i}^{(\alpha_p\rho_p)}$ and $\theta_{i}^{(\alpha_p)}$ are the positivity limiters based on $(\alpha_p\rho_p)$ and $\alpha_p$, respectively, and $(\theta_p)_{i} = \min\left(\theta_{i}^{(\alpha_p\rho_p)}, \theta_{i}^{(\alpha)_p}\right)$. The formulation of $\theta_{i}^{(f)}$ \citep{Zhang2017,HuangJohnsen2024} is
\begin{equation}\label{Eq:Limiter-LowerBound}
\theta_{i}^{(f)}=\min\left(1,\frac{\overline{f}_i-f_{\min}}{\overline{f}_i-\min\left(f_i^C,f_{i+1/2}^L,f_{i-1/2}^R\right)}\right) \in [0,1],
\end{equation}
where $f_{i}^C=\left(\overline{f}_i-\hat{\omega}^{(1)} f_{i-1/2}^R - \hat{\omega}^{(N_{quad})} f_{i+1/2}^L\right)/\left(1-\hat{\omega}^{(1)}-\hat{\omega}^{(N_{quad})}\right)$ from the conservation requirement (see Eq.~(\ref{Eq:CellAverage})).
The usage of $(\theta_p)_{i}$ for both $(\alpha_p\rho_p)_{i+1/2}^L$ and $(\alpha_p)_{i+1/2}^L$ in Eq.~(\ref{Eq:WENO-C-Positivity}) is to continuously satisfy the consistency requirement.

The third step is to preserve volume fraction summation to unity, i.e., $\sum_{p=1}^N (\alpha_p)_{i+1/2}^L=\sum_{p=1}^N (\alpha_p)_{i-1/2}^R=1$ \citep{HuangJohnsen2023,HuangJohnsen2024}. We first calculate $\mathcal{S}_{i+1/2}^L = \sum_{p=1}^N (\alpha_p)_{i+1/2}^L$, and then update $(\alpha_p\rho_p)_{i+1/2}^L$ and $(\alpha_p)_{i+1/2}^L$ from
\begin{equation}\label{Eq:WENO-C-Summation}
(\alpha_p\rho_p)_{i+1/2}^L \leftarrow \frac{(\alpha_p\rho_p)_{i+1/2}^L}{\mathcal{S}_{i+1/2}^L},
\quad
(\alpha_p)_{i+1/2}^L \leftarrow \frac{(\alpha_p)_{i+1/2}^L}{\mathcal{S}_{i+1/2}^L}.
\end{equation}
As $(\alpha_p\rho_p)_{i+1/2}^L$ and $(\alpha_p)_{i+1/2}^L$ are positive before Eq.~(\ref{Eq:WENO-C-Summation}) is implemented, their updated values are still positive. Moreover, the updated $(\alpha_p)_{i+1/2}^L$ satisfies $\sum_{p=1}^N (\alpha_p)_{i+1/2}^L=1$, and thus $(\alpha_p)_{i+1/2}^L$ (for all $p$) is bounded in $[0,1]$. The consistency requirement is again satisfied as $(\alpha_p\rho_p)_{i+1/2}^L$ and $(\alpha_p)_{i+1/2}^L$ are updated in the same manner in Eq.~(\ref{Eq:WENO-C-Summation}). It should be noted that $\sum_{p=1}^N (\alpha_p)_{i}^C=1$ is also satisfied because $(\alpha_p)_i^C$ is linearly related to $(\alpha_p)_{i+1/2}^L$, $(\alpha_p)_{i-1/2}^R$, and $\overline{(\alpha_p)}_i$ from the conservation requirement (Eq.~(\ref{Eq:CellAverage})).

After the third step, the reconstructed masses and volume fractions at cell faces are admissible (mass positivity, volume fraction boundedness, and volume fraction summation to unity in Eq.~(\ref{Eq:AdmissibleSet})). However, to satisfy the admissibility (bound preservation) requirement \citep{HuangJohnsen2024}, the reconstructed data inside grid cells ($(\alpha_p\rho_p)_i^C$ and $(\alpha_p)_i^C$) must be admissible as well. As $\sum_{p=1}^N (\alpha_p)_i^C=1$ is satisfied, the only task is to ensure the positivity of $(\alpha_p\rho_p)_i^C$ and $(\alpha_p)_i^C$, which is achieved by applying the positivity limiter in Eq.~(\ref{Eq:Limiter-LowerBound}) again. Due to the conservation requirement, $(\alpha_p\rho_p)_{i+1/2}^L$ and $(\alpha_p)_{i+1/2}^L$ are updated accordingly via
\begin{equation}\label{Eq:WENO-C-Admissibility}
(\alpha_p\rho_p)_{i+1/2}^L \leftarrow \overline{(\alpha_p\rho_p)}_i + \theta_i \left( (\alpha_p\rho_p)_{i+1/2}^L-\overline{(\alpha_p\rho_p)}_i\right),
\quad
(\alpha_p)_{i+1/2}^L \leftarrow \overline{(\alpha_p)}_i + \theta_{i} \left( (\alpha_p)_{i+1/2}^L-\overline{(\alpha_p)}_i \right),
\end{equation}
where $\theta_{i} = \min_p( (\theta_p)_i)$ and $(\theta_p)_i$ is recalculated with the updated $(\alpha_p\rho_p)$ and $\alpha_p$ from Eq.~(\ref{Eq:WENO-C-Summation}).
To preserve the consistency property as well as volume-fraction summation to unity, the same limiter value needs to be applied to both $(\alpha_p\rho_p)_{i+1/2}^L$ and $(\alpha_p)_{i+1/2}^L$ for all the phases. As a result, to achieve the positivity of $(\alpha_p\rho_p)_i^C$ and $(\alpha_p)_i^C$ for all the phases, $\theta_{i}$, the minimum value of $\{(\theta_p)_i\}_{p=1}^N$, is used.
Now, the reconstructed masses and volume fractions are admissible (see Eq.~(\ref{Eq:AdmissibleSet})).

After finishing the above procedure, the reconstructed primitive variables $\mathbf{V}_{i+1/2}^L$ and $\mathbf{V}_{i-1/2}^R$ are obtained, resulting in $\mathbf{U}_{i+1/2}^L=\psi_{\mathbf{V} \rightarrow \mathbf{U}}(\mathbf{V}_{i+1/2}^L)$ and $\mathbf{U}_{i-1/2}^R=\psi_{\mathbf{V} \rightarrow \mathbf{U}}(\mathbf{V}_{i-1/2}^R)$. From the conservation requirement, $\mathbf{U}_{i}^C$ is obtained from $\mathbf{U}_{i+1/2}^L$, $\mathbf{U}_{i-1/2}^R$, and $\overline{\mathbf{U}}_{i}$ by inverting Eq.~(\ref{Eq:CellAverage}).
As the masses and volume fractions are not changed by the mapping $\psi_{\mathbf{V} \rightarrow \mathbf{U}}(\cdot)$, they are still admissible. The final step is to enforce the remaining energy constraint in Eq.~(\ref{Eq:AdmissibleSet}) via
\begin{equation}\label{Eq:WENO-C-Energy}
\mathbf{U}_{i+1/2}^L \leftarrow \overline{\mathbf{U}}_i + \theta_i^{(\psi)} \left(\mathbf{U}_{i+1/2}^L-\overline{\mathbf{U}}_i\right),
\end{equation}
where $\theta_i^{(\psi)}$ is the positive limiter based on $\psi$ in Eq.~(\ref{Eq:EnergyBound}).
We note that, in the calculation of $\theta_i^{(\psi)}$ from Eq.~(\ref{Eq:Limiter-LowerBound}), $\overline{\psi}_i=\psi(\overline{\mathbf{U}}_i)$, $\psi_{i+1/2}^L=\psi(\mathbf{U}_{i+1/2}^L)$, $\psi_{i-1/2}^R=\psi(\mathbf{U}_{i-1/2}^R)$, and $\psi_{i}^C=\psi(\mathbf{U}_{i}^C)$. As analyzed in \citep{Zhang2017,HuangJohnsen2024}, the updated $\mathbf{U}_{i+1/2}^L$ and $\mathbf{U}_{i-1/2}^R$ from Eq.~(\ref{Eq:WENO-C-Energy}) and the resulting $\mathbf{U}_{i}^C$ from inverting Eq.~(\ref{Eq:CellAverage}) all satisfy the energy constraint in Eq.~(\ref{Eq:AdmissibleSet}). Moreover, since a single value of $\theta_i^{(\psi)}$ is applied to all components of $\mathbf{U}$, the masses and volume fractions are still admissible. As a result, the final $\mathbf{U}_{i+1/2}^L$, $\mathbf{U}_{i-1/2}^R$, and $\mathbf{U}_{i}^C$ are in $\mathcal{U}$ (admissibility or bound preservation), and satisfy Eq.~(\ref{Eq:CellAverage}) (the conservation requirement). For the same reason, the consistency and equilibrium properties are not affected. Therefore, the proposed reconstruction scheme satisfies all the requirements in Section~\ref{Sec:Scheme-Requirement}, as reflected by $(\alpha_1\rho_1)$ and the temperature error (labeled by ``WENO-C'') shown in Fig.~\ref{Fig:Advection}.
Furthermore, the proposed approach is robust; it does not need to compute the phasic density $\rho_p=(\alpha_p\rho_p)/\alpha_p$, which is not well-defined as $\alpha_p$ approaches $0$.
Hereafter, ``WENO-C'' is used to represent the proposed WENO reconstruction scheme.

The consistency of reduction \citep{BoyerMinjeaud2014,Dong2018,Huangetal2020N,Huangetal2020B} is built into the WENO-C scheme; if a phase is absent in the reconstruction stencil, its reconstructed mass and volume fraction are both zero from Eq.~(\ref{Eq:WENO-C-MassVolume}), Eq.~(\ref{Eq:WENO-C-Positivity}), Eq.~(\ref{Eq:WENO-C-Summation}), Eq.~(\ref{Eq:WENO-C-Admissibility}), and Eq.~(\ref{Eq:WENO-C-Energy}). As a result, no fictitious phase is produced, and the reconstruction for the other phases is not affected.

From a theoretical perspective, it is possible to develop an alternative positivity limiter to ensure only the positivity of $f_{i+1/2}^{L}$ and $f_{i-1/2}^{R}$ in Eq.~(\ref{Eq:WENO-C-Positivity}), and another one that ensures only the positivity of $f_i^C$ in Eq.~(\ref{Eq:WENO-C-Admissibility}) ($f=(\alpha_p\rho_p)$, $\alpha_p$). However, in practice, it is more convenient to have a single formulation for the positivity limiter throughout the reconstruction, and, therefore, we use that in Eq.~(\ref{Eq:Limiter-LowerBound}), which satisfies both the requirements for Eq.~(\ref{Eq:WENO-C-Positivity}) and Eq.~(\ref{Eq:WENO-C-Admissibility}).

In terms of accuracy, the WENO-C scheme has the same order of accuracy as the WENO-W scheme implemented for $(\alpha_p\rho_p)$ and $\alpha_p$ in Eq.~(\ref{Eq:WENO-C-MassVolume}) and for $\mathbf{u}$ and $P$. As analyzed and demonstrated in \citep{ZhangShu2010,ZhangShu2010Euler,Zhang2017}, the positivity limiter implemented in Eq.~(\ref{Eq:WENO-C-Positivity}) and Eq.~(\ref{Eq:WENO-C-Energy}) does not influence the order of accuracy in smooth regions. The operations in Eq.~(\ref{Eq:WENO-C-Summation}) and Eq.~(\ref{Eq:WENO-C-Admissibility}) are effective only in interfacial regions, where the order of accuracy reduces to (at best) first order \citep{LeVeque2002}.

The reconstructed polynomials of $\mathbf{U}$ from the WENO-C scheme can be obtained following the approach in \citep{ZhangShu2010}, although, in practice, there is no need to store the polynomial coefficients. If the fifth-order WENO-C scheme is used, with $\overline{\mathbf{U}}_{i-1}$, $\overline{\mathbf{U}}_{i}$, $\overline{\mathbf{U}}_{i+1}$, and the final $\mathbf{U}_{i+1/2}^L$ and $\mathbf{U}_{i-1/2}^R$ from Eq.~(\ref{Eq:WENO-C-Energy}), the reconstructed polynomials in cell $I_i$ are
\begin{equation}\label{Eq:WENO-C-Polynomial}
\begin{split}
\mathbf{U}_i(x) &= \mathcal{B}_0 + \mathcal{B}_1 (x - x_i) + \mathcal{B}_2 (x - x_i)^2 + \mathcal{B}_3 (x - x_i)^3 + \mathcal{B}_4 (x - x_i)^4,\\
\mathcal{B}_0 &= \frac{1}{192} \left( \left(\overline{\mathbf{U}}_{i+1} + 298 \overline{\mathbf{U}}_i + \overline{\mathbf{U}}_{i-1}\right) - 54 \left(\mathbf{U}_{i+1/2}^L + \mathbf{U}_{i-1/2}^R\right) \right),\\
\mathcal{B}_1 &= \frac{-1}{8 \Delta x} \left( \left(\overline{\mathbf{U}}_{i+1} - \overline{\mathbf{U}}_{i-1}\right) + 10\left(\mathbf{U}_{i-1/2}^R - \mathbf{U}_{i+1/2}^L\right) \right),\\
\mathcal{B}_2 &= \frac{-1}{8 \Delta x^2} \left( \left( \overline{\mathbf{U}}_{i+1} + 58 \overline{\mathbf{U}}_{i} + \overline{\mathbf{U}}_{i-1}\right) - 30\left(\mathbf{U}_{i+1/2}^L + \mathbf{U}_{i-1/2}^R\right) \right),\\
\mathcal{B}_3 &= \frac{1}{2 \Delta x^3} \left( \left(\overline{\mathbf{U}}_{i+1} - \overline{\mathbf{U}}_{i-1}\right) + 2 \left(\mathbf{U}_{i-1/2}^R - \mathbf{U}_{i+1/2}^L\right) \right),\\
\mathcal{B}_4 &= \frac{5}{12 \Delta x^4} \left( \left(\overline{\mathbf{U}}_{i+1} + 10\overline{\mathbf{U}}_{i} + \overline{\mathbf{U}}_{i-1}\right) - 6\left(\mathbf{U}_{i+1/2}^L + \mathbf{U}_{i-1/2}^R\right) \right),
\end{split}   
\end{equation}
for uniform mesh size $\Delta x$.

The proposed WENO-C scheme is described generally for an arbitrary number of phases, while, for two-phase flows, it is common that only $\alpha_1$ is stored and solved. To implement the WENO-C scheme in two-phase flows, we start with temporarily obtaining $\overline{(\alpha_2)}=1-\overline{(\alpha_1)}$ in the reconstruction stencil. Then, the WENO-C scheme is implemented exactly following the procedure described here.

\subsubsection{Variant with the consistent limiter}\label{Section:Limiter-Consistent}
We further propose a variant of WENO-C, which alternatively uses the consistent limiter \citep{HuangJohnsen2023} to enforce volume fraction summation to unity. Specifically, Eq.~(\ref{Eq:WENO-C-Summation}) of the above procedure is replaced with
\begin{equation}\label{Eq:WENO-C-Summation-Limiter}
(\alpha_p\rho_p)_{i+1/2}^L \leftarrow \overline{(\alpha_p\rho_p)}_{i}+\theta_{i+1/2}^{(\alpha_p)} S_{i+1/2}^{(\alpha_p\rho_p)} \frac{\Delta x_{i}}{2},
\qquad
(\alpha_p)_{i+1/2}^L \leftarrow \overline{(\alpha_p)}_{i}+\theta_{i+1/2}^{(\alpha_p)} S_{i+1/2}^{(\alpha_p)} \frac{\Delta x_{i}}{2},
\end{equation}
where $S_{i+1/2}^{(f)}$ is an approximation of the gradient of $f$ at $x_{i+1/2}$ in cell $I_i$ obtained from
\begin{equation}\label{Eq:Slope}
S_{i+1/2}^{(\alpha_p\rho_p)}
=
\frac{(\alpha_p\rho_p)_{i+1/2}^L-\overline{(\alpha_p\rho_p)}_{i}}{\Delta x_{i}/2},
\qquad
S_{i+1/2}^{(\alpha_p)}
=
\frac{(\alpha_p)_{i+1/2}^L-\overline{(\alpha_p)}_{i}}{\Delta x_{i}/2},
\end{equation}
and $\theta_{i+1/2}^{(\alpha_p)}$ is the consistent limiter \citep{HuangJohnsen2023} obtained from
\begin{equation}\label{Eq:Limiter-Consistent}
\theta_{i+1/2}^{(\alpha_p)}=
\left\{
\begin{array}{cc}
\max\left( 
\min\left(
\frac{ \hat{S}_{i+1/2}^{(\alpha_{\hat{p}})} }{ S_{i+1/2}^{(\alpha_{\hat{p}})} },1
\right),
0 
\right), & p=\hat{p},\\
\max\left( 
\min\left( 
\frac{ S_{i+1/2}^{(\alpha_{\hat{p}})} }{ \hat{S}_{i+1/2}^{(\alpha_{\hat{p}})} } ,1 
\right),
0 
\right), & \mathrm{else},
\end{array}
\right.
\end{equation}    
with $\hat{S}_{i+1/2}^{(\alpha_{\hat{p}})}=S_{i+1/2}^{(\alpha_{\hat{p}})} - \sum_{p=1}^N S_{i+1/2}^{(\alpha_p)}$ and $\hat{p} = \arg\max_p |S_{i+1/2}^{(\alpha_p)}|$. This variant of the WENO-C scheme with the consistent limiter is termed ``WENO-CL''. We note that the implementation of the consistent limiter to high-order schemes here is simpler than that in \citep{HuangJohnsen2023,HuangJohnsen2024}; there is no need to limit individual Taylor series coefficients of the reconstructed polynomials of $(\alpha_p\rho_p)$ and $\alpha_p$.

As shown in \citep{HuangJohnsen2023}, the consistent limiter satisfies $\sum_{p=1}^N \theta_{i+1/2}^{(\alpha_p)} S_{i+1/2}^{(\alpha_p)}=0$ and $\theta_{i+1/2}^{(\alpha_p)} \in [0,1]$. Thanks to these two properties, it is clear from Eq.~(\ref{Eq:WENO-C-Summation-Limiter}) that the updated $\{(\alpha_p)_{i+1/2}^L\}_{p=1}^N$ satisfy $\sum_{p=1}^N (\alpha_p)_{i+1/2}^L=1$ due to $\sum_{p=1}^N \theta_{i+1/2}^{(\alpha_p)} S_{i+1/2}^{(\alpha_p)}=0$, and the positivity of $(\alpha_p\rho_p)_{i+1/2}^L$ and $(\alpha_p)_{i+1/2}^L$ is preserved due to $\theta_{i+1/2}^{(\alpha_p)} \in [0,1]$, resulting in $(\alpha_p)_{i+1/2}^L \in [0,1]$.
Furthermore, the consistency requirement is satisfied as $\theta_{i+1/2}^{(\alpha_p)}$ is also applied to $(\alpha_p\rho_p)$.
Finally, consistency of reduction is also satisfied because $S_{i+1/2}^{(\alpha_p)}$ becomes zero when $\alpha_p$ is absent in the reconstruction stencil.
Again, these operations are in effect only in interfacial regions.

Different from WENO-C that modifies all non-zero volume fractions to enforce volume fraction summation to unity (see Eq.~(\ref{Eq:WENO-C-Summation})), WENO-CL achieves the same property without modifying the volume fractions for which the gradient is zero, thanks to the consistent limiter. Therefore, WENO-CL is favored in cases that allow for a uniform value of volume fraction other than $0$ or $1$ to appear, e.g., inert species in combustion \citep{BaumgartBlanquart2024}. However, it is impossible for a volume fraction to have a non-zero and non-unity constant value when the phases are immiscible, and thus WENO-C, or more specifically, Eq.~(\ref{Eq:WENO-C-Summation}), is feasible in this scenario. The problems of interest in the present study are immiscible multiphase flows, and thus WENO-C is majorly investigated because it is simpler and more efficient, while the properties of WENO-CL are also verified.

\subsubsection{Implementation with adaptive mesh refinement}\label{Sec:AMR}
The proposed WENO-C (as well as WENO-CL) scheme can be readily incorporated into the block-structured adaptive mesh refinement (AMR), following the framework of \citep{Huangetal2025} for compressible multiphase flows with the Phase-Field mechanism. Specifically, AMR consists of multiple levels of meshes that have different grid sizes, while each level of the mesh is uniform. The proposed WENO-C (as well as WENO-CL) scheme can thus be directly applied on individual levels of the uniform meshes without any modification.

\section{Results}\label{Sec:Results}
Fifth-order WENO schemes and third-order TVD Runge-Kutta time stepping \citep{Shu1988,GottliebShu1998} are used for both the hyperbolic (Eq.~(\ref{Eq:Hyperbolic})) and Phase-Field (Eq.~(\ref{Eq:PhaseField})) steps.
As a specific case of Eq.~(\ref{Eq:EOS-Bound-General}), the equation of state by Le M{\'e}tayer et al. \citep{LeMetayeretal2005}
\begin{equation}\label{Eq:EOS}
\mathsf{e}_p^{(1)}(P)=A_p^P P +B_p^P,\quad
\mathsf{e}_p^{(2)}(P)=D_p,\quad
\mathsf{e}_p^T(T_p,\rho_p)=\frac{\rho_p C_p^T T_p + B_p^T+D_p\rho_p}{\rho_p},\quad
1 \leqslant p \leqslant N,
\end{equation}
is used, where $\gamma_p$, $P_p^\infty$, $C_p^T$, and $D_p$ are material properties, and $A_p^P=\frac{1}{\gamma_p-1}$, $B_p^P=\frac{\gamma_p P_p^\infty}{\gamma_p-1}$, and $B_p^T=P_p^\infty$. The pressure is calculated from
\begin{equation}\label{Eq:Pressure}
P
=
\frac{1}{\sum_{p=1}^N A_p^P \alpha_p}\left(
(\rho E)-\frac{1}{2}\frac{(\rho \mathbf{u})\cdot(\rho \mathbf{u})}{\sum_{p=1}^N(\alpha_p\rho_p)}
-
\sum_{p=1}^N B_p^P \alpha_p
-
\sum_{p=1}^N D_p (\alpha_p \rho_p)
\right).
\end{equation}

We calculate $\widehat{[\nabla \cdot {\mathbf{J}}_p]}$ in Eq.~(\ref{Eq:Q-E}) with second-order central differences for the Laplacian and the mid-point rule for the integrals in the multiphase reduction-consistent conservative Allen-Cahn model \citep{Huangetal2020B}, which reads
\begin{eqnarray}\label{Eq:CAC}
[\nabla \cdot \mathbf{J}_p]
=
M \nabla^2 \phi_p 
- 
\frac{M}{\eta^2} \left( g'(\phi_p)-\phi_p \sum_{q=1}^N g'(\phi_q) \right)
+
L_p^c,
\quad
g(\phi)=\phi^2 (1-\phi)^2,
\quad
L_p^c=\sum_{q=1}^N W_{p,q} B_q,\\
\nonumber
\sum_{q=1}^N \left( \int_\Omega W_{p,q} d\Omega \right) B_q
=
\int_\Omega \frac{M}{\eta^2} \left( g'(\phi_p)-\phi_p \sum_{q=1}^N g'(\phi_q) \right) d\Omega,
\quad
W_{p,q}=\left\{\begin{array}{cc}
     -\phi_p \phi_q,&  p\neq q,\\
     \phi_p (1-\phi_q),&  p=q,
\end{array}
\right.
\end{eqnarray}
where $M=0.5 \eta \max(\sqrt{\mathbf{u}\cdot\mathbf{u}})$ is the mobility and $\eta/\Delta x=1$ controls the interface thickness, as in \citep{HuangJohnsen2022,HuangJohnsen2023,HuangJohnsen2024,Huangetal2025}. Then, the numerical Phase-Field flux vector proposed in \citep{HuangJohnsen2024} is implemented, which preserves admissibility (physical bounds) as explained in Section~\ref{Sec:Scheme-Requirement}.

\subsection{Verification problems}\label{Sec:Verification}
We first present results verifying the analysis of the WENO schemes in Section~\ref{Sec:WENO}. The HLLC approximate Riemann solver \citep{Toroetal1994,Toro2009} is used to obtain the numerical hyperbolic flux vector. When the thermal equilibrium is investigated, the temperature is calculated from
\begin{equation}\label{Eq:Temperature}
T
=
\frac{1}{\sum_{p=1}^N C_p^T (\alpha_p\rho_p)}\left(
(\rho E)-\frac{1}{2}\frac{(\rho \mathbf{u})\cdot(\rho \mathbf{u})}{\sum_{p=1}^N(\alpha_p\rho_p)}
-
\sum_{p=1}^N B_p^T \alpha_p
-
\sum_{p=1}^N D_p (\alpha_p \rho_p)
\right).
\end{equation}
The interface thickness of a phase (or the number of grid cells across its interface) is estimated from its volume fraction by \citep{HuangJohnsen2022,HuangJohnsen2023,HuangJohnsen2024,Huangetal2025}
\begin{equation}\label{Eq:InterfaceThickness}
N_I[\alpha]=\frac{\int_{\Omega} \mathcal{H}(\alpha) d\Omega}{\Delta x \int_{\Omega} |\nabla \alpha|d\Omega},
\qquad
\mathcal{H}(\alpha)=\left\{
\begin{array}{cc}
     1,&  \alpha_{\min} \leqslant \alpha \leqslant \alpha_{\max},\\
     0,&  \mathrm{else},
\end{array}
\right.
\end{equation}
where $\alpha_{\min}=0.05$ and $\alpha_{\max}=0.95$.

\subsubsection{Two-phase advection}\label{Sec:Advection-TwoPhase}
To compare the performance of WENO-JS, WENO-W, and WENO-C schemes in compressible multiphase flows, we consider a two-phase advection problem with air and water in \citep{HuangJohnsen2022}.
The doubly periodic domain is $[-0.5,0.5]\times[-0.5,0.5]$, discretized with $100\times100$ grid cells. An air circle (Phase 1: $\rho_{1}=1.204\times10^{-3}$, $\gamma_1=1.4$, $P_{1}^{\infty}=0$, and $D_{1}=0$) surrounded by water (Phase 2: $\rho_{2}=1$, $\gamma_2=6.12$, $P_{2}^{\infty}=0.1631$, and $D_{2}=0$) is initially at $(x_r,y_r)=(0,0)$ with a radius of $r=0.2$. Both the air and water are translated by a uniform velocity $\mathbf{u}_0=(1,1)$ with a uniform pressure $P_0=4.819\times10^{-5}$ and temperature $T_0=300$. To achieve thermal equilibrium, $\{C_p^T\}_{p=1}^2$ are obtained from $C_p^T=(A_p^P P_0 + B_p^P - B_p^T)/(\rho_p T_0)$.
The time step is fixed to be $\Delta t=1\times10^{-3}$, while the Phase-Field step is not activated.

To continue the comparison following Fig.~\ref{Fig:Advection}, Fig.~\ref{Fig:Advection-TwoPhase-Contour} shows the contours of $(\alpha_1\rho_1)$ and $\alpha_1$ at $t=1$ (after one period of advection) with MUSCL (minmod limiter) \citep{LeVeque2002}, WENO-JS, WENO-W, and WENO-C schemes. We include the MUSCL scheme because it is consistent by construction. As shown by these contours, the WENO schemes have weaker numerical diffusion than the MUSCL scheme, resulting in a slower increase of the interface thickness. However, both Fig.~\ref{Fig:Advection} and Fig.~\ref{Fig:Advection-TwoPhase-Contour} illustrate two types of oscillation produced by the WENO-JS scheme; obvious oscillations in $(\alpha_1\rho_1)$ near the air-water interface and small oscillations in both $(\alpha_1\rho_1)$ and $\alpha_1$ in the bulk-phase regions. These oscillations are not observed with the MUSCL, WENO-W, and WENO-C schemes, thus demonstrating the effectiveness of the proposed modified WENO weight in Eq.~(\ref{Eq:WENO-W-Weight}). 
\begin{figure}[!t]
	\centering
	\includegraphics[scale=.25]{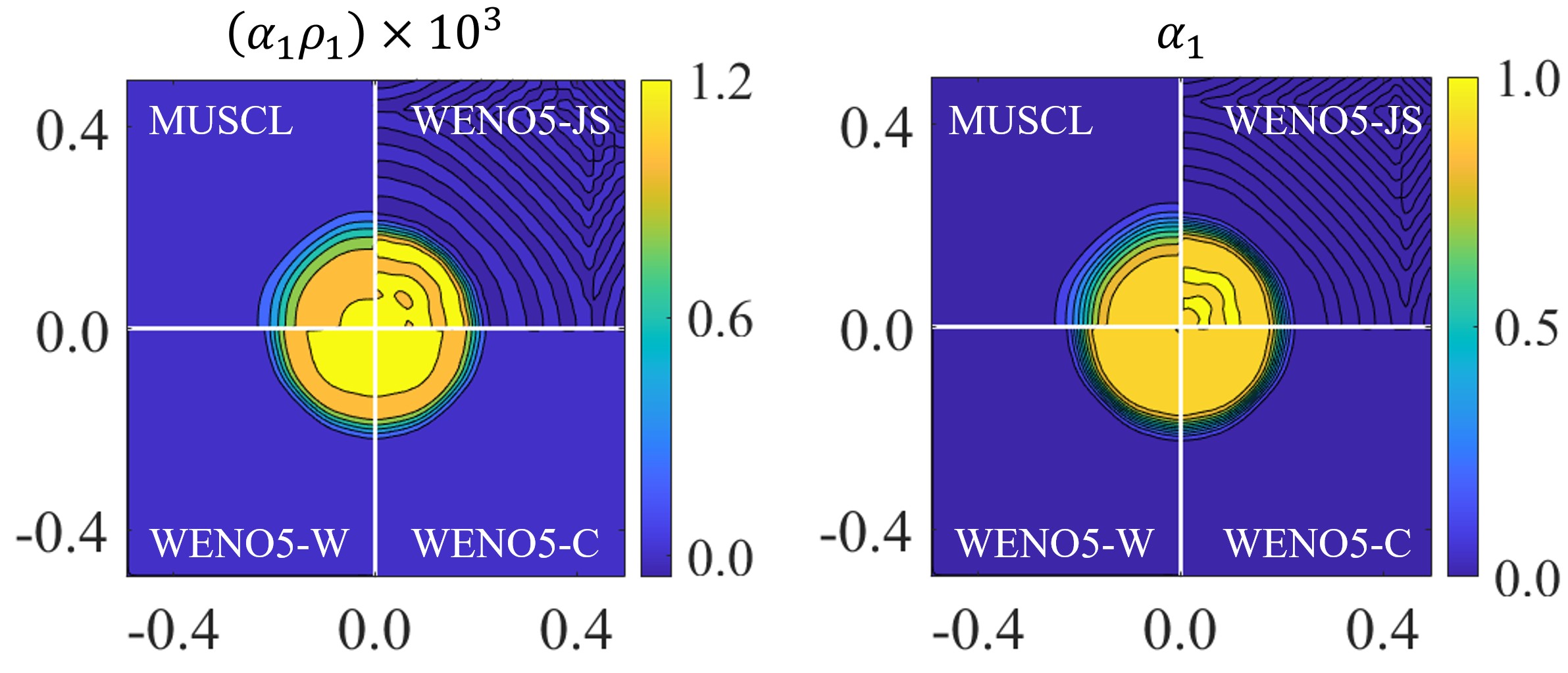}
	\caption{Contours of $(\alpha_1\rho_1)\times10^{3}$ (left) and $\alpha_1$ (right) at $t=1$ in the two-phase advection problem with the MUSCL, WENO5-JS, WENO5-W, and WENO5-C schemes.}\label{Fig:Advection-TwoPhase-Contour}
\end{figure}

\begin{figure}[!t]
	\centering
	\includegraphics[scale=.25]{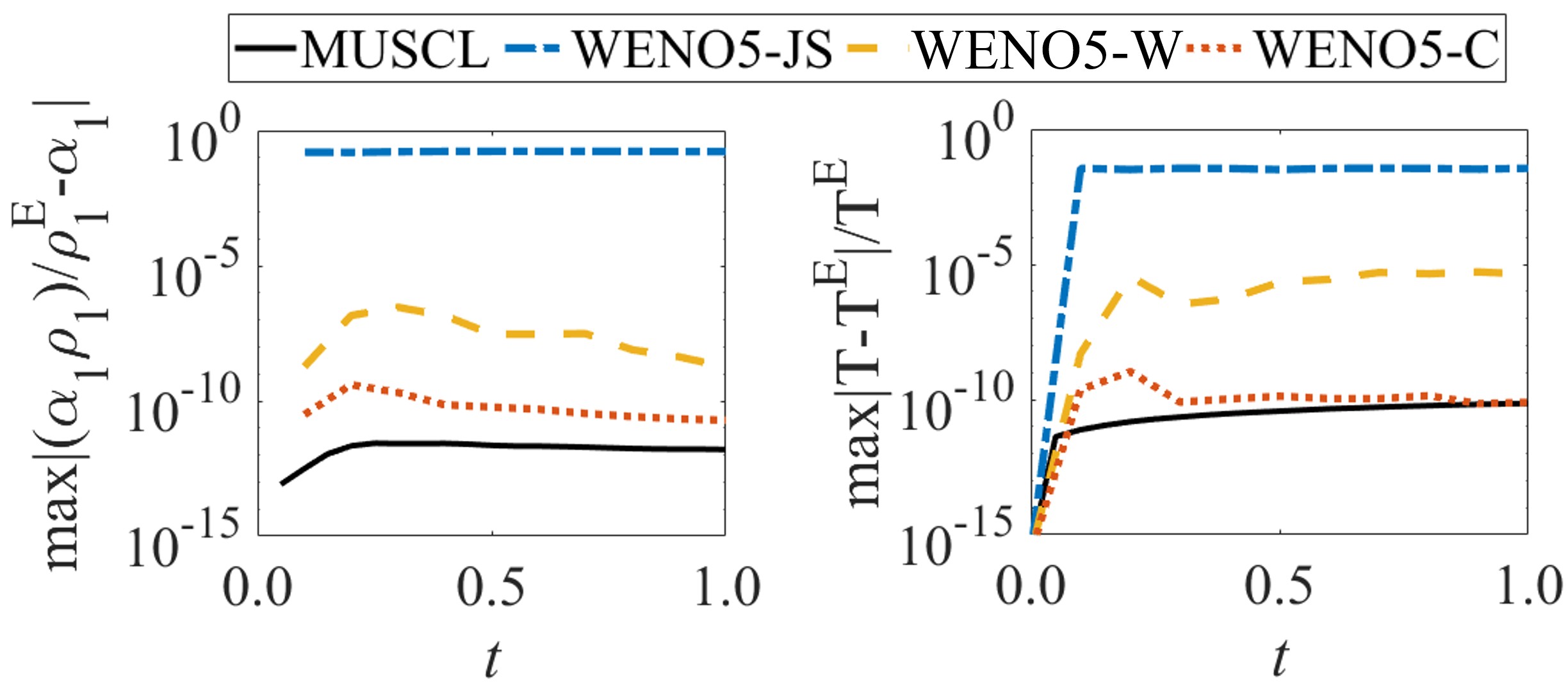}
	\caption{Time histories of the consistency (left) and temperature (right) errors in the two-phase advection problem with the MUSCL, WENO5-JS, WENO5-W, and WENO5-C schemes.}\label{Fig:Advection-TwoPhase-Error}
\end{figure}

Fig.~\ref{Fig:Advection-TwoPhase-Error} further shows the time histories of the consistency error ($(\alpha_1\rho_1)/\rho_1^E-\alpha_1$) and the temperature error $((T-T^E)/T^E)$. The consistency error and the corresponding temperature error have the same order of magnitude, because the temperature error is actually produced by the consistency error, as analyzed in \citep{HuangJohnsen2022,HuangJohnsen2023,HuangJohnsen2024}. Since the WENO-JS scheme fails the consistency requirement (see Section~\ref{Sec:WENO-JS}), it produces the most significant error in ($(\alpha_1\rho_1)/\rho_1^E-\alpha_1$), which is reflected in the temperature error as well. Although the WENO-W scheme is consistent theoretically and reduces the consistency error by about nine orders of magnitude in practice, the effect of the round-off error limits its performance, as reflected by the temperature error.
This issue is further alleviated by the WENO-C scheme; both the consistency and temperature errors are further reduced to magnitudes similar to those of the MUSCL scheme.
Based on this comparison, we only implement the proposed WENO-C scheme in the following sections.

\subsubsection{Five-phase advection}\label{Sec:Advection-FivePhase}
We consider the five-phase advection problem \citep{HuangJohnsen2023} to further verify the properties of the proposed WENO-C scheme in problems with more than two phases. 
The doubly periodic domain is $[-0.5,0.5]\times[-0.5,0.5]$, discretized with $100\times100$ grid cells. The time step is fixed to be $\Delta t=1\times10^{-3}$. The initial uniform velocity and pressure are $\mathbf{u}_0=(1,1)$ and $P_0=4.819\times10^{-5}$, respectively. The material properties and the initial locations of the phases are listed in Table~\ref{Table:Advection-FivePhase}.
\begin{table}[]
    \centering
    \begin{tabular}{c|c|c|c|c|cc}
    \hline
         Phase index&                $\rho$&      $\gamma$&   $P^{\infty}$&                     $D$&           $(x_r,y_r,r)$&\\
         \hline
                   1&  $1.204\times10^{-3}$&       $1.400$&        $0.000$&   $0.000$              &              Background&\\
                   2&  $1.000$             &       $6.120$&        $0.163$&   $0.000$              &       $(0.00,0.00,0.20)$&\\        
                   3&  $0.800$             &       $2.350$&        $0.476$&   $-5.550\times10^{-4}$&       $(0.00,\pm0.50,0.15)$&\\
                   4&  $0.100\times10^{-3}$&       $1.600$&        $0.000$&   $0.000$              &       $(\pm0.50,0.00,0.15)$&\\
                   5&  $1.000\times10^{-3}$&       $3.000$&        $0.000$&   $0.000$              &                   Absent&\\
    \hline
    \end{tabular}
    \caption{Material properties and initial locations of the phases in the five-phase advection problem.}
    \label{Table:Advection-FivePhase}
\end{table}

\begin{figure}[!t]
	\centering
    \includegraphics[scale=0.4]{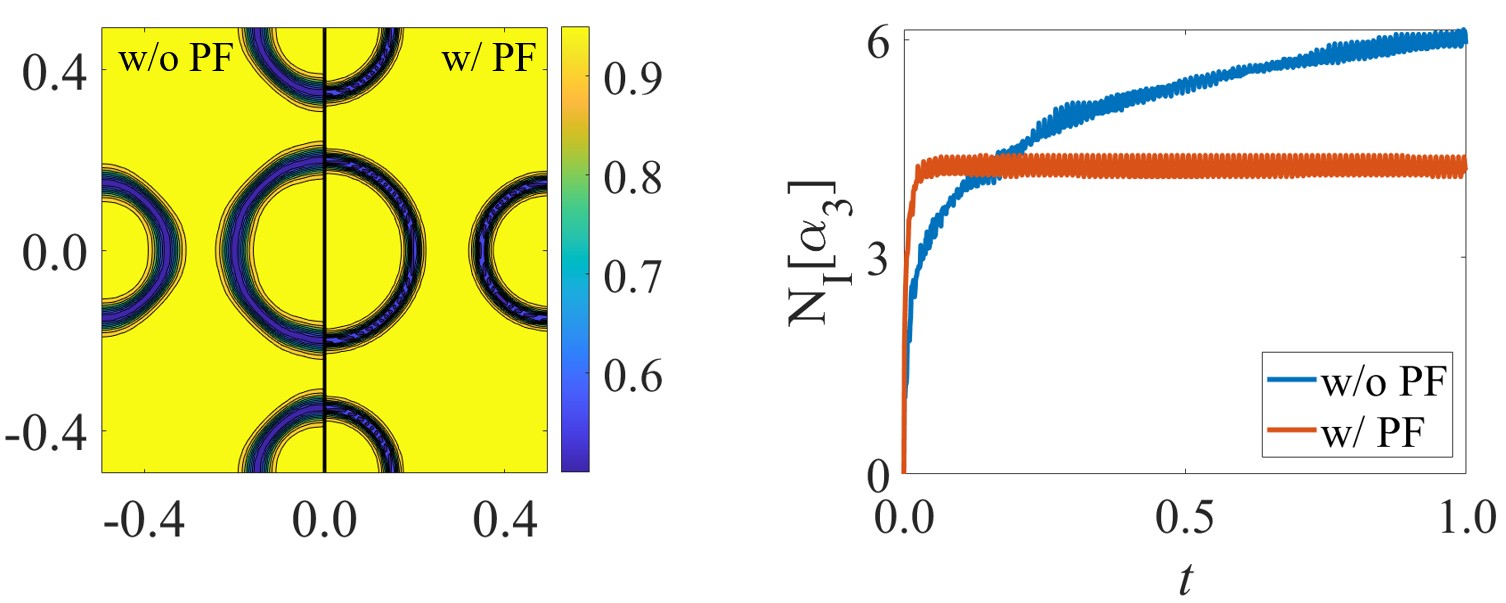}
	\caption{Contour of $\sum_{p=1}^N\alpha_p^2$ at $t=1$ (left) and time history of the interface thickness (right) with and without the Phase-Field mechanism in the five-phase advection problem.}\label{Fig:Advection-FivePhase-Contour}
\end{figure}

Fig.~\ref{Fig:Advection-FivePhase-Contour} shows the contour of $\sum_{p=1}^N \alpha_p^2$ at $t=1$, along with the time history of the interface thickness of Phase~$3$. $\sum_{p=1}^N \alpha_p^2$ highlights the interfacial regions where the volume fractions are neither $0$ nor $1$. The interfaces are sharper and spread more uniformly for all the phases with the Phase-Field mechanism, which is further quantified by the time history of the interface thickness. Using Eq.~(\ref{Eq:InterfaceThickness}), we observe a monotonically increasing interface thickness without the Phase-Field mechanism, while, with the Phase-Field mechanism, a fixed interface thickness of $4$ to $5$ grid cells is achieved after an initial sharp rise.

\begin{figure}[!t]
	\centering
    \includegraphics[scale=0.35]{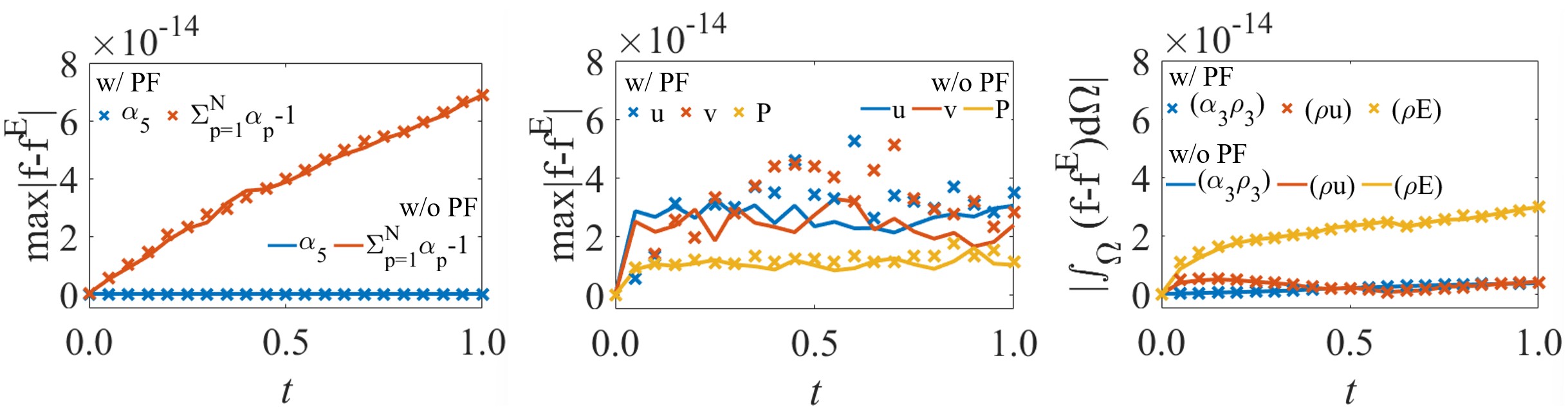}
	\caption{Time histories of the volume fraction (left), equilibrium (middle), and conservation errors (right) with and without the Phase-Field mechanism in the five-phase advection problem.}\label{Fig:Advection-FivePhase-Error}
\end{figure}

Fig.~\ref{Fig:Advection-FivePhase-Error} shows the time histories of the volume fraction, equilibrium, and conservation errors; all the errors are of the order of round-off error, regardless of whether the Phase-Field mechanism is activated or not. The consistency and temperature errors behave similarly to those in Fig.~\ref{Fig:Advection-TwoPhase-Error}, and are not repeated here. Therefore, it is verified that the proposed WENO-C scheme satisfies volume fraction summation to unity ($\sum_{p=1}^N \alpha_p=1$), consistency of reduction ($\alpha_5=0$ given $\alpha_5|_{t=0}=0$), equilibrium of velocity, pressure, and temperature, and conservation of mass, momentum, and energy, simultaneously, even having more than two phases.

\subsubsection{Four-phase advection with non-zero and non-unity constant volume fractions}\label{Sec:Advection-Miscible}
To verify the properties of the WENO-CL scheme in Section~\ref{Section:Limiter-Consistent}, the multiphase advection problems are modified such that $\alpha_1|_{t=0}=0.8$ inside the circle $(x_r,y_r,r)=(0.0,0.0,0.2)$, while $\alpha_2|_{t=0}=0.8$ outside, and $\alpha_3|_{t=0}=0.2$ and $\alpha_4|_{t=0}=0$ (absent) in the entire domain. Such an initial phase configuration is not allowed in immiscible multiphase flows, but can appear in problems like combustion. The material properties of the four phases are the same as those listed in Table~\ref{Table:Advection-FivePhase}.

\begin{figure}[!t]
	\centering
        \includegraphics[scale=0.33]{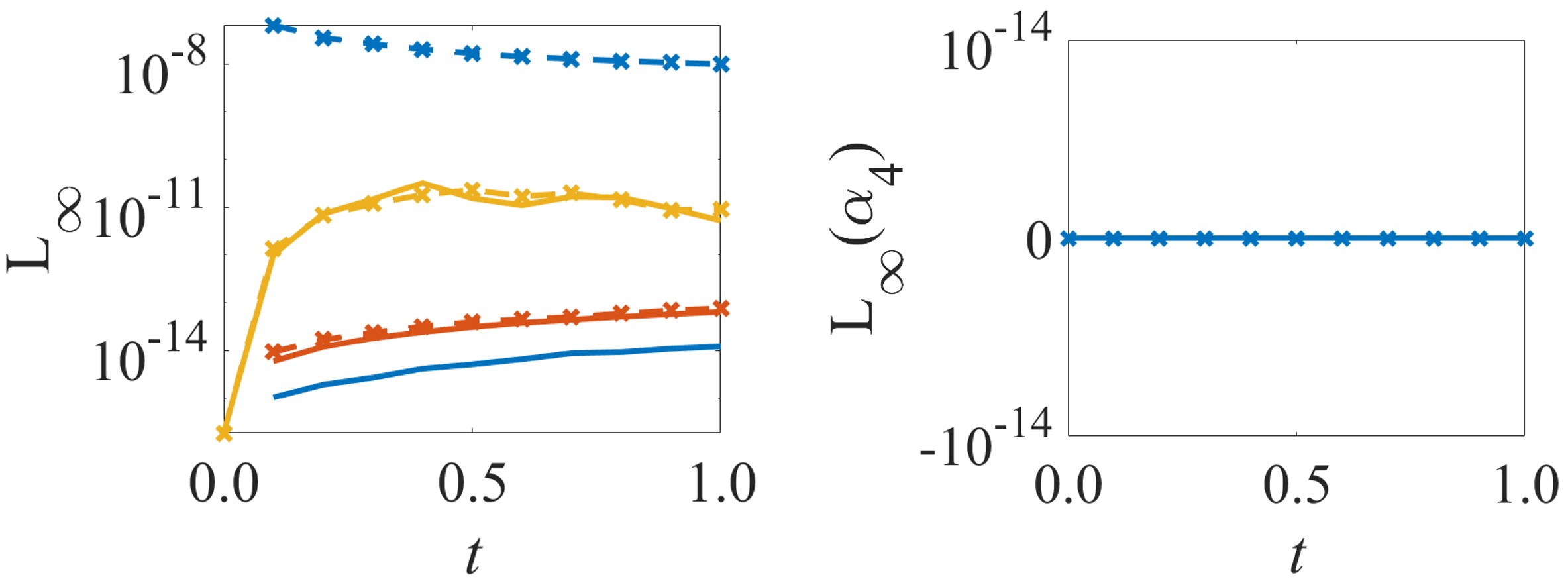}
	\caption{Time histories of the errors in the four-phase advection problem with non-zero and non-unity constant volume fractions. Left: $\alpha_3$ (blue), $\sum_{p=1}^N \alpha_p$ (red), and temperature (orange); Right: $\alpha_4$; Solid line: WENO5-CL, Dashed line with cross: WENO5-C.}\label{Fig:WENO-CL}
\end{figure}

Fig.~\ref{Fig:WENO-CL} shows the time histories of the errors in temperature, volume fraction summation, $\alpha_3$, and $\alpha_4$, with both WENO-CL (solid line) and WENO-C (dashed line with cross). As expected, the WENO-CL scheme maintains $\alpha_3=0.2$ up to the round-off error, thanks to the consistent limiter \citep{HuangJohnsen2023}, while the WENO-C scheme produces a significant error of $\alpha_3$ due to Eq.~(\ref{Eq:WENO-C-Summation}). This kind of $\alpha_3$ error is also observed in \citep{BaumgartBlanquart2024}, where WENO-JS was used.
Other than that, volume fraction summation to unity, consistency between mass and volume fraction, and consistency of reduction are all satisfied by both WENO-C and WENO-CL, as reflected by the errors in $\sum_{p=1}^N \alpha_p$, temperature, and $\alpha_4$, respectively. This test case again verifies that the usage of WENO-C, or more specifically Eq.~(\ref{Eq:WENO-C-Summation}) to enforce volume fraction summation to unity, should be limited to immiscible multiphase problems.

\subsubsection{Two-phase advection with a smooth volume fraction}\label{Sec:Advection-Smooth}
To verify the order of accuracy of the proposed WENO-C scheme, we consider an advection problem with a smooth initial volume fraction $\alpha_1|_{t=0}=0.2 \times ( 3+\sin( 2 \pi x ) )$. The initial uniform velocity and pressure are $u_0=1$ and $P_0=1$, respectively. The material properties are $(\rho_1,\gamma_1,P_1^{\infty},D_1)=(2,1.5,0,0)$ and $(\rho_2,\gamma_2,P_2^{\infty},D_2)=(1,1.01,0.02,3)$. The periodic domain is $[0,1]$. The time step is determined by $\Delta t=(\Delta x)^{5/3}$, as the time stepping is third-order accurate.

\begin{figure}[!t]
	\centering
	\includegraphics[scale=0.3]{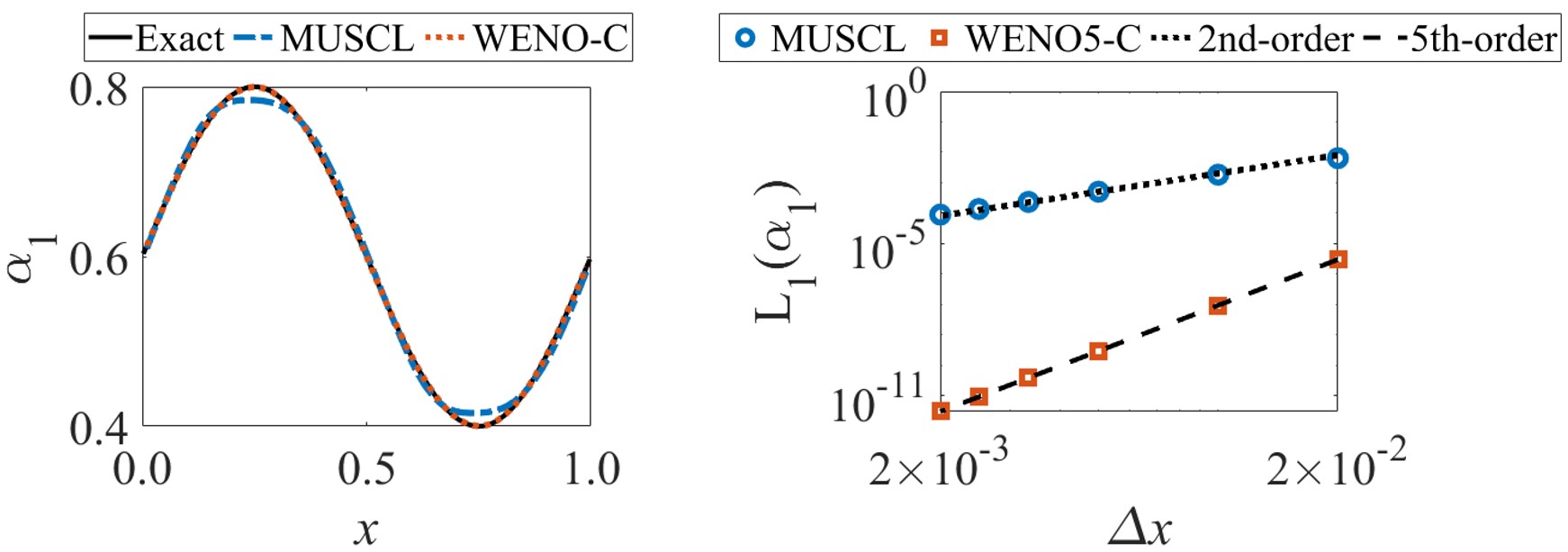}
	\caption{$\alpha_1$ with $\Delta x=0.02$ (left) and $L_1$ error of $\alpha_1$ (right) at $t=1$ in the two-phase advection problem with a smooth volume fraction.}\label{Fig:Advection-Smooth}
\end{figure}

Fig.~\ref{Fig:Advection-Smooth} shows $\alpha_1$ with $\Delta x=0.02$ and the $L_1$ error of $\alpha_1$ with respect to the cell size at $t=1$. Both the fifth-order WENO-C scheme and the MUSCL scheme (with the minmod limiter) are used for comparison. For this smooth problem, as expected, the WENO-C scheme outperforms the MUSCL scheme in particular near the smooth extrema. Furthermore, the $L_1$ error of $\alpha_1$ verifies that the WENO-C scheme is fifth-order accurate, while the MUSCL scheme is second-order accurate.
    
\subsubsection{Air-helium shock tube}\label{Sec:AirHeliumShock}
To verify the order of accuracy of the proposed WENO-C scheme in problems with both shocks and interfaces, we consider an air-helium shock tube problem with initial conditions:
\begin{equation}\label{Eq:AirHeliumShock-IC}
(\rho_1,\rho_2,u,P,\alpha_1,\alpha_2)=\left\{
\begin{array}{cc}
  (0.125,1,0,1,0,1),   &  0 < x \leqslant 1,\\
  (0.125,1,0,0.1,1,0),   &  1 < x < 2.
\end{array}
\right.
\end{equation}
Both the air and helium are modeled as ideal gases with $\gamma_1=1.6$ (helium) and $\gamma_2=1.4$ (air). The CFL number is $CFL=0.4$.

\begin{figure}[!t]
	\centering
	\includegraphics[scale=0.3]{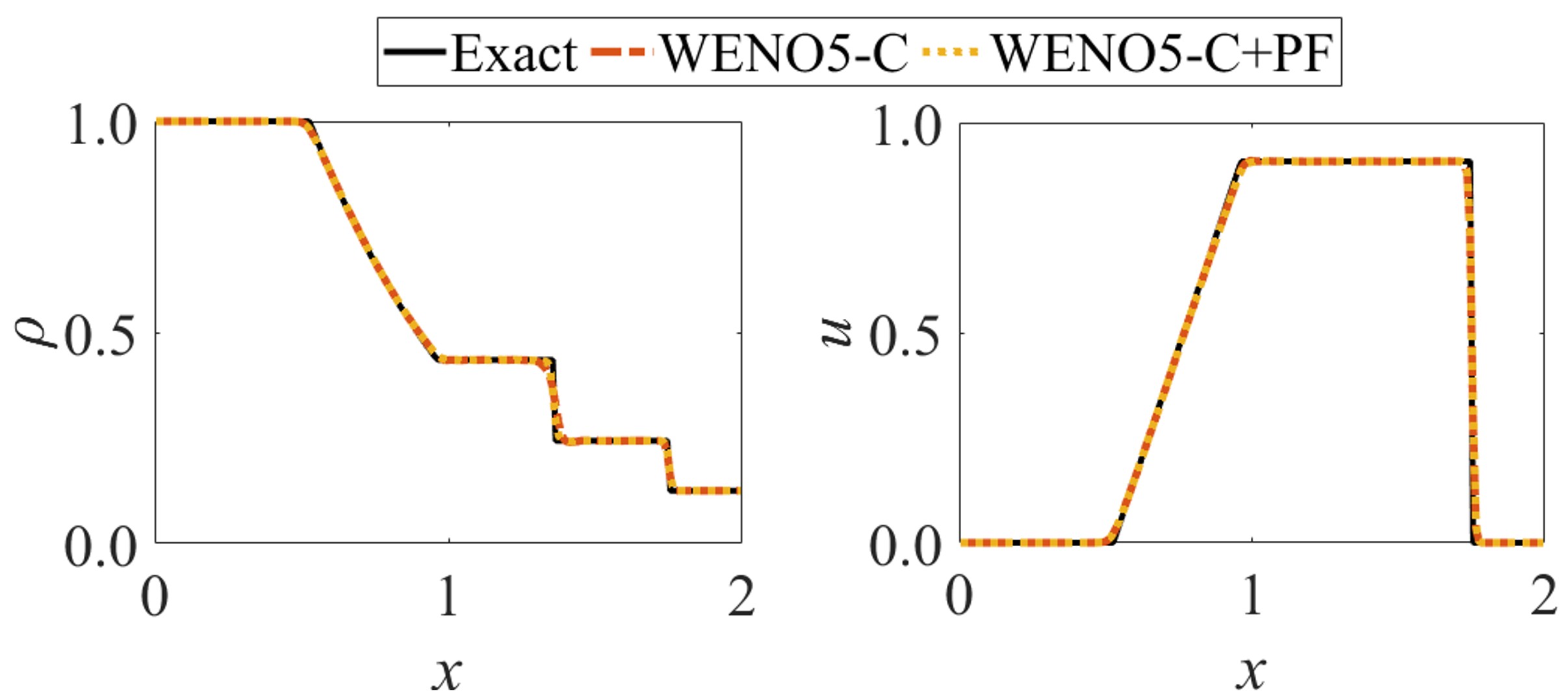}\\
    \includegraphics[scale=0.3]{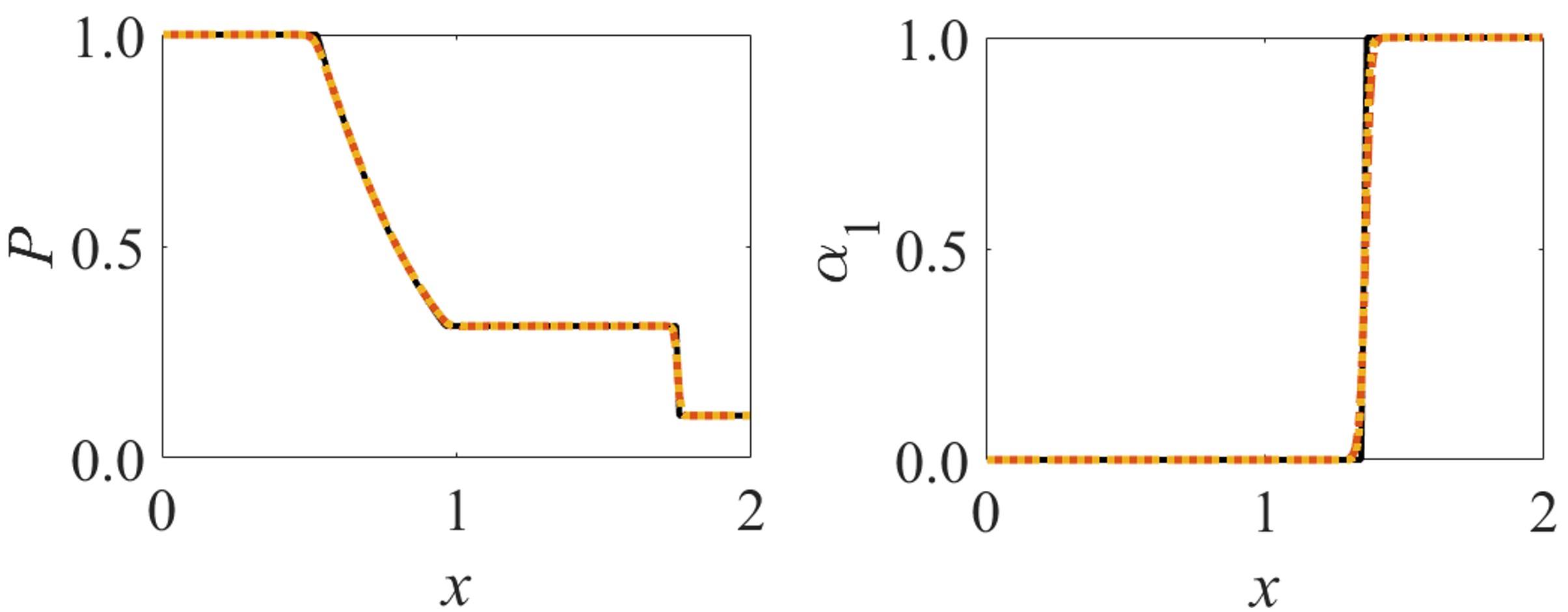}
	\caption{Density (top left), velocity (top right), pressure (bottom left), and volume fraction (bottom right) at $t=0.4$ with $\Delta x=0.01$ in the air-helium shock tube problem.}\label{Fig:AirHelium}
\end{figure}

\begin{figure}[!t]
	\centering
	\includegraphics[scale=0.3]{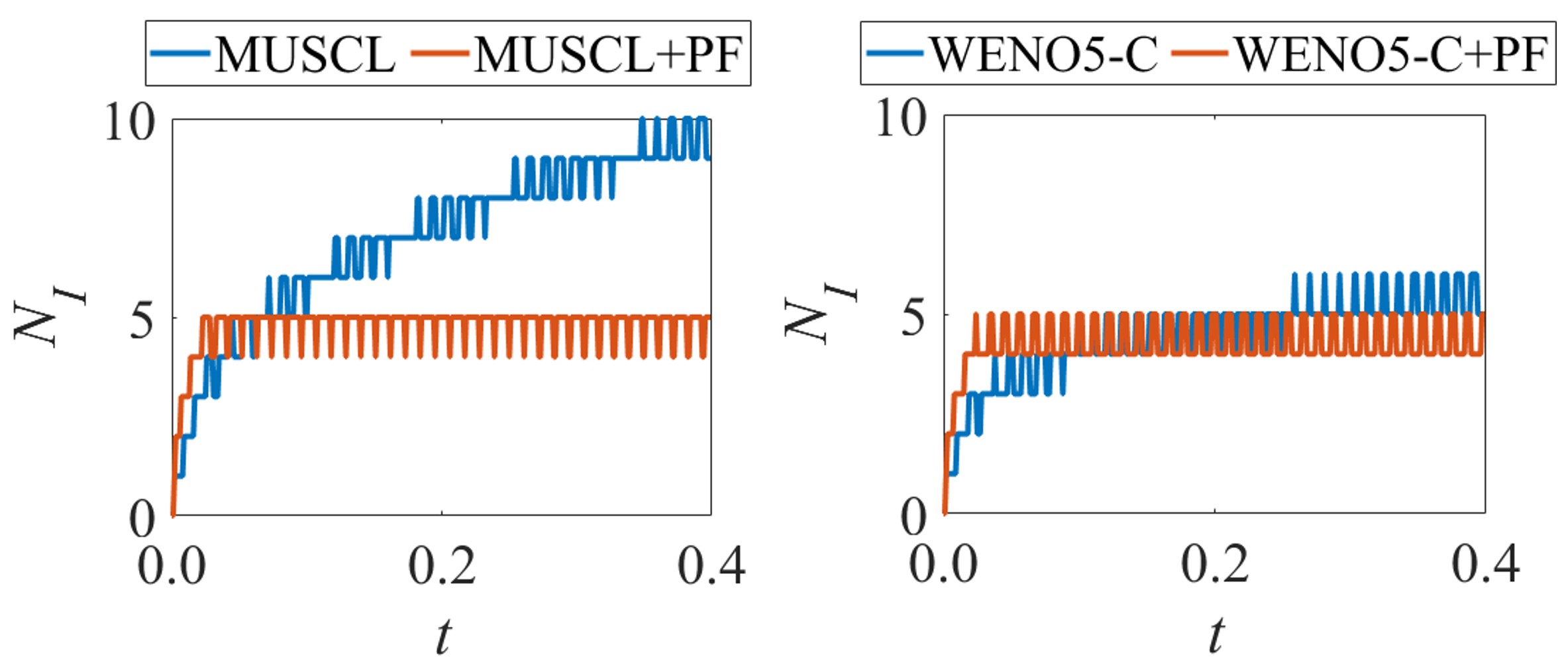}
	\caption{Time history of the interface thickness with the MUSCL (left) and WENO5-C (right) schemes in the air-helium shock tube problem.}\label{Fig:AirHelium-Thickness}
\end{figure}

\begin{figure}[!t]
	\centering
	\includegraphics[scale=0.3]{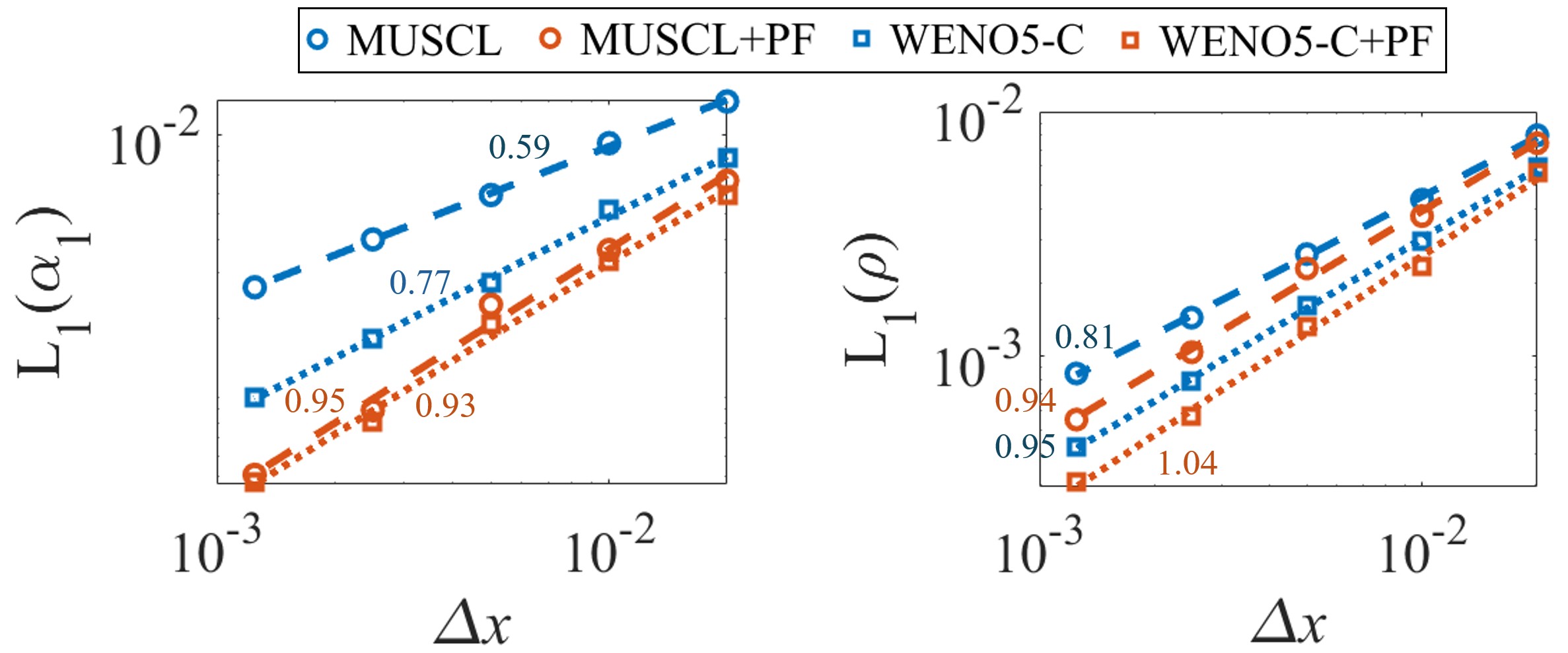}
	\caption{$L_1$ errors of the volume fraction (left) and the density (right) with respect to the cell size in the air-helium shock tube problem.}\label{Fig:AirHelium-L1}
\end{figure}

Fig.~\ref{Fig:AirHelium} shows the density, velocity, pressure, and volume fraction at $t=1$ with $\Delta x=0.01$. Both the results with and without the Phase-Field mechanism agree well with the exact solution. Fig.~\ref{Fig:AirHelium-Thickness} and Fig.~\ref{Fig:AirHelium-L1} further quantify the effect of the Phase-Field mechanism. 

Fig.~\ref{Fig:AirHelium-Thickness} shows the time history of the interface thickness ($N_I$). As expected, when the Phase-Field mechanism is not activated, the interface thickness continuously increases over time, though at a much higher rate with MUSCL than with fifth-order WENO-C. After the Phase-Field mechanism is included, the interface thickness is fixed to be $4$ to $5$ grid cells, the same as that in our previous study \citep{HuangJohnsen2022,HuangJohnsen2023,HuangJohnsen2024}, regardless of whether the WENO-C or MUSCL scheme is used.

Fig.~\ref{Fig:AirHelium-L1} shows and the $L_1$ errors of the volume fraction and the density with respect to the cell size, including the actual convergence rates fitted from the $L_1$ errors. In general, the fifth-order WENO-C scheme has smaller errors and faster convergence rates than the MUSCL scheme, and the scheme with the Phase-Field mechanism outperforms the corresponding scheme without it. As this problem has both shocks and a material interface, the optimal convergence rate is $1$st-order \citep{LeVeque2002}, which is achieved only when the Phase-Field mechanism is activated.
For completeness, the accuracy tests in this section and Section~\ref{Sec:Advection-Smooth} are repeated for WENO-CL, and similar behaviors are observed, see \ref{Appendix:WENO-CL-Accuracy}.

\subsection{Application problems}\label{Sec:Application}
\begin{figure}[!t]
	\centering
	\includegraphics[scale=0.33]{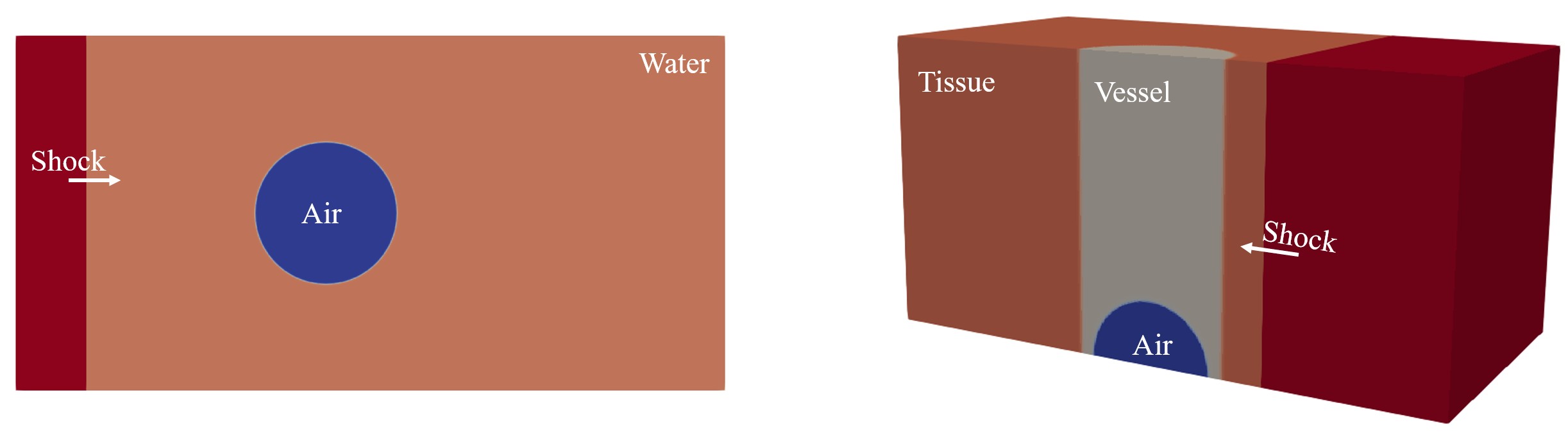}
	\caption{Schematic of the shock-induced collapse of an air cavity in water (left) and the shock--vessel--bubble interaction (right).}\label{Fig:Application-IC}
\end{figure}

We further implement the proposed WENO-C scheme in two challenging compressible multiphase flows: the shock-induced collapse of an air cavity in water \citep{Shuklaetal2010} (two-dimensional, two-phase) and the shock--vessel--bubble interaction \citep{CoralicColonius2014} (three-dimensional, three-phase), which are both illustrated in Fig.~\ref{Fig:Application-IC}. These two problems were previously investigated with the models of Allaire et al. \citep{Allaireetal2002}, although it has been shown in \citep{Tiwarietal2013,Schmidmayeretal2020} that the model of Allaire et al. \citep{Allaireetal2002} ($K_p=0$) failed in spherical bubble collapse but the model of Kapila et al. \citep{Kapilaetal2001} ($K_p=(\rho c^2)/(\rho_p c_p^2)-1$) succeeded. Therefore, in our study, we use both the models of Allaire et al. \citep{Allaireetal2002} and Kapila et al. \citep{Kapilaetal2001} with the Phase-Field mechanism to, on one hand, demonstrate the success of WENO-C in solving different compressible multiphase flow models and, on the other hand, further illustrate their differences in bubble collapse dynamics.
Preserving the admissibility (physical bounds) plays a crucial role in successfully simulating these challenging problems, and thus the HLL flux \citep{HuangJohnsen2024,Toro2009} is used in the hyperbolic step. Following the analysis in \citep{HuangJohnsen2024}, the corresponding sufficient CFL condition for bound preservation is $CFL \leqslant \hat{\omega}^{(1)}=1/12$, because the fifth-order WENO-C scheme requires a fourth-order polynomial in each grid cell (see Eq.~(\ref{Eq:WENO-C-Polynomial})), resulting in $N_{quad}=4$ and $\hat{\omega}^{(1)}=\hat{\omega}^{(N_{quad})}=1/12$ for the Gauss-Lobatto quadrature rule. However, our preliminary tests show that these two problems can be run with $CFL=0.2$, as the sufficient condition is usually more restrictive than the necessary condition.

\subsubsection{Shock-induced collapse of an air cavity in water}\label{Sec:Cavity}
An air cavity (Phase~$1$: $\gamma_1 = 1.4$, $P^{\infty}_{1} = 0.0$, and $D_1=0.0$) in water (Phase~$2$: $\gamma_2 = 4.4$, $P^{\infty}_2 = 6000.0$, and $D_2=0.0$) is impinged upon by a Mach $1.72$ shock in the water \citep{Shuklaetal2010}.
The initial conditions for this two-dimensional problem are
\begin{equation}\label{Eq:Cavity-IC}
(\rho_1,\rho_2,u,v,P,\alpha_1)=\left\{
\begin{array}{cc}
(1.0\times10^{-3},1.000, 0.00, 0.0,1.000\times10^{0} ,1),&  \psi \geqslant 0,\\
(1.0\times10^{-3},1.000, 0.00, 0.0,1.000\times10^{0} ,0),&  \psi < 0 \quad \& \quad x \geqslant x_S,\\
(1.0\times10^{-3},1.325,68.52, 0.0,1.915\times10^{4} ,0),&  \psi < 0 \quad \& \quad x < x_S,
\end{array}
\right.
\end{equation}
where $x_S=1.0$ and $\psi=r - \sqrt{ (x-x_r)^2 + (y-y_r)^2 }$ with $x_r=4.375$, $y_r=0.0$, and $r=1.0$, as shown in Fig.~\ref{Fig:Application-IC} (left). The domain is $[0.0,10.0]\times[-2.5,2.5]$ with outflow boundary conditions and is discretized with $1600\times800$ grid cells.

\begin{figure}[!t]
	\centering
	\includegraphics[scale=0.4]{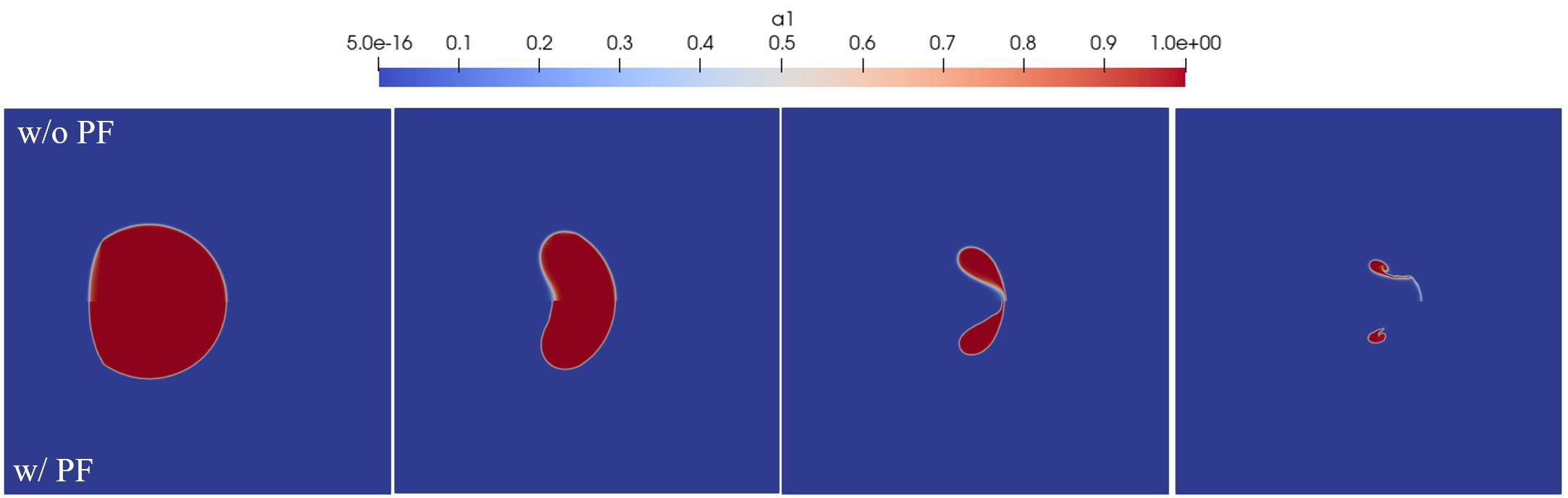}\\
    \includegraphics[scale=0.4]{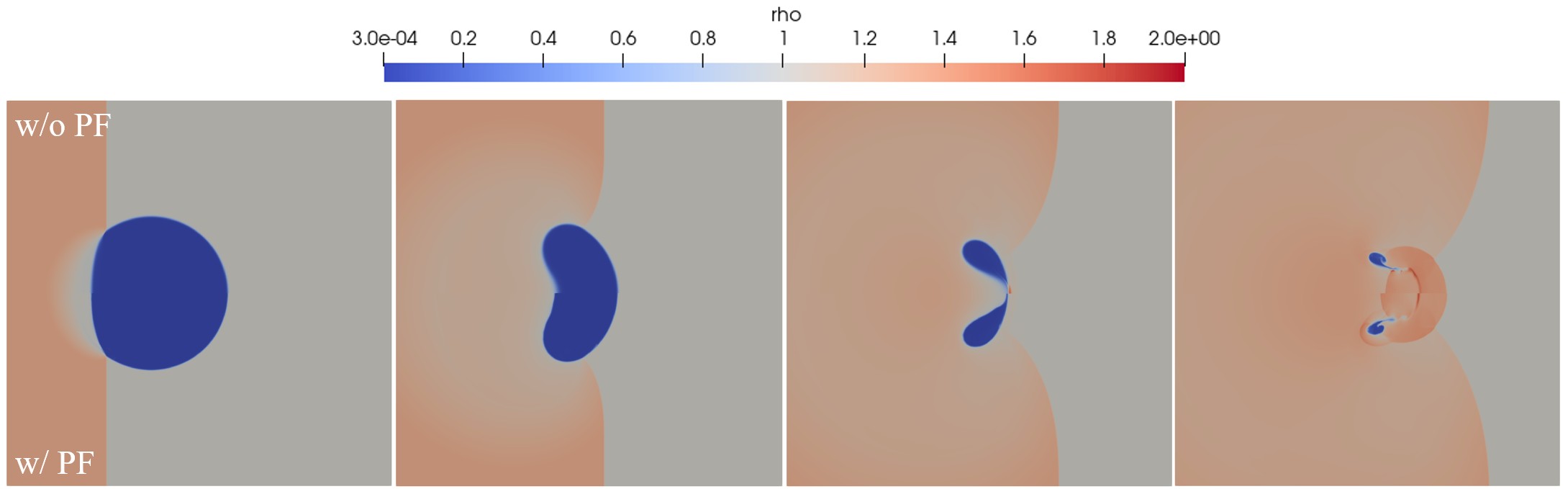}\\
    \includegraphics[scale=0.4]{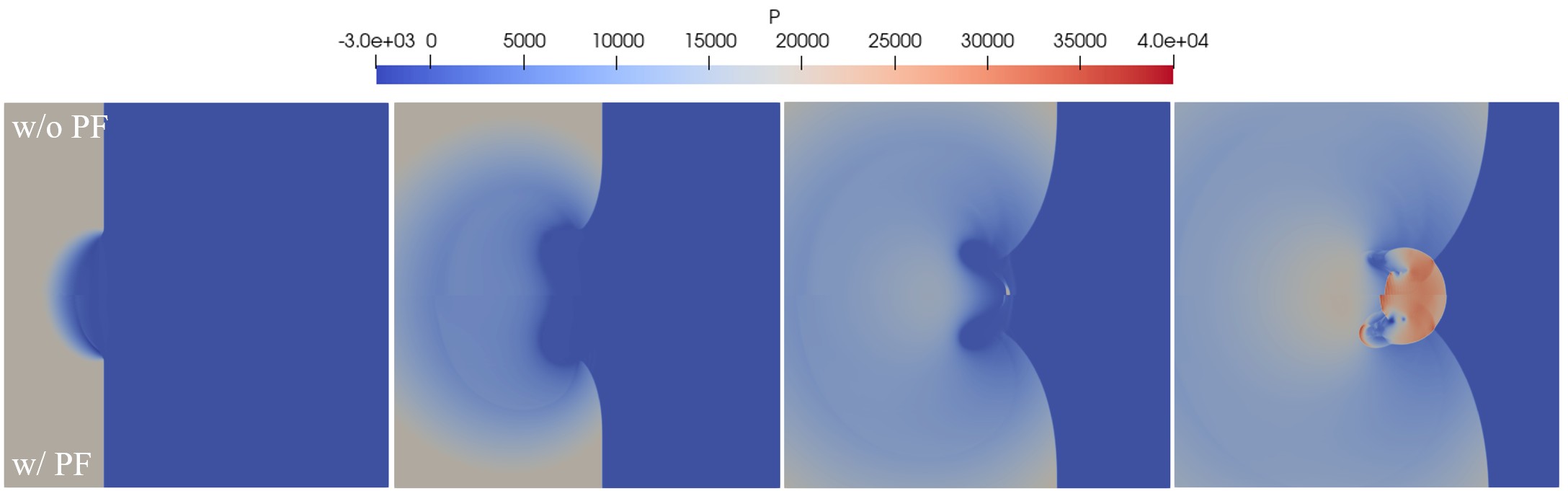}
	\caption{Volume fraction (top), density (middle), and pressure (bottom) in the shock-induced collapse of an air cavity in water at $t=0.010$, $0.015$, $0.018$, and $0.20$ (from left to right) without (top frame) and with (bottom frame) the Phase-Field mechanism with the model of Allaire et al. \citep{Allaireetal2002}.}\label{Fig:Cavity-A}
\end{figure}

\begin{figure}[!t]
	\centering
    \includegraphics[scale=0.4]{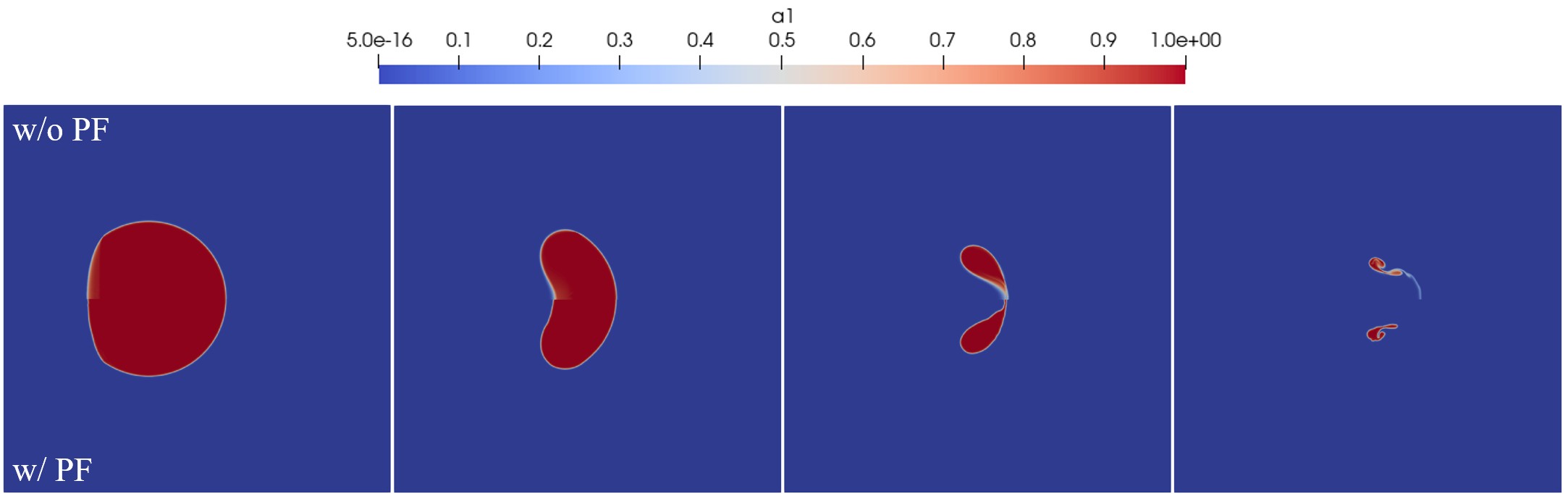}\\
    \includegraphics[scale=0.4]{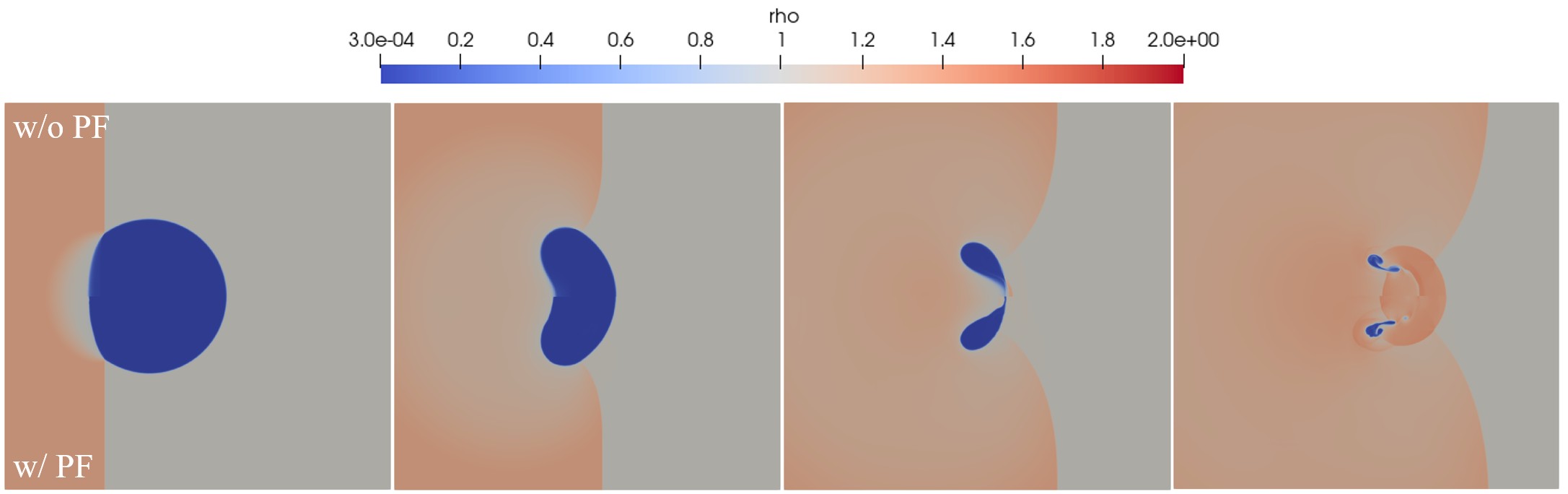}\\
    \includegraphics[scale=0.4]{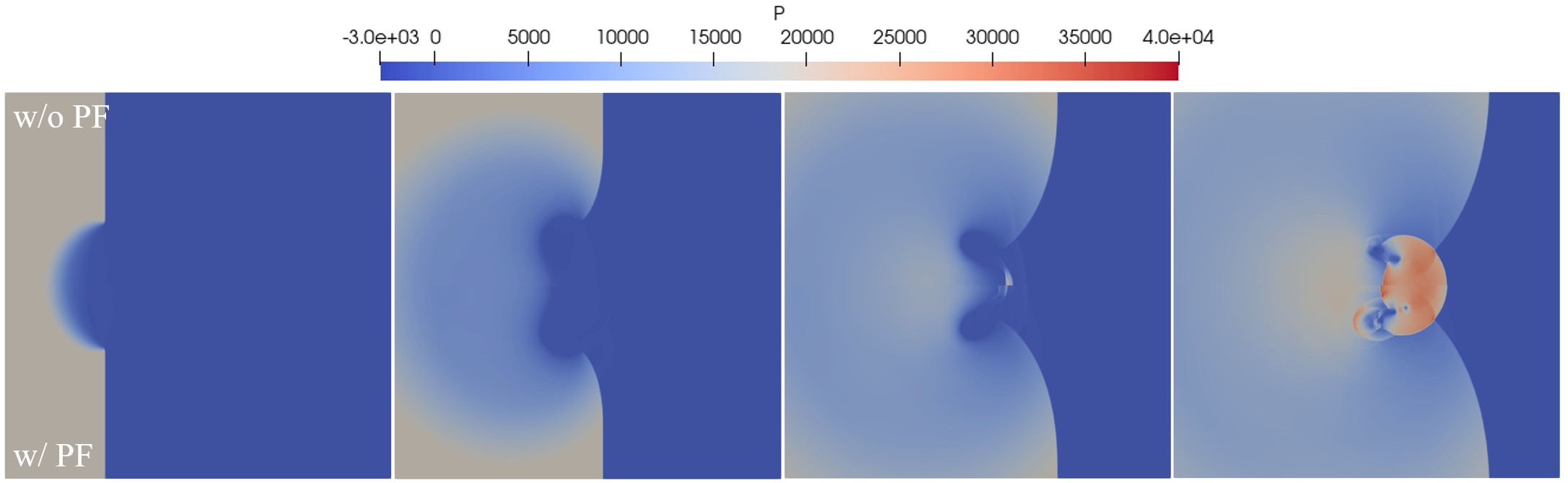}
	\caption{Volume fraction (top), density (middle), and pressure (bottom) in the shock-induced collapse of an air cavity in water at $t=0.010$, $0.015$, $0.018$, and $0.20$ (from left to right) without (top frame) and with (bottom frame) the Phase-Field mechanism with the model of Kapila et al. \citep{Kapilaetal2001}.}\label{Fig:Cavity-K}
\end{figure}

Fig.~\ref{Fig:Cavity-A} and Fig.~\ref{Fig:Cavity-K} show the volume fraction, density, and pressure at selected instants without and with the Phase-Field mechanism using the models of Allaire et al. \citep{Allaireetal2002} and Kapila et al. \citep{Kapilaetal2001}, respectively. As the shock interacts with the bubble, a re-entrant jet forms on the proximal side, penetrating the bubble as it collapses. Eventually, the jet impinges upon the distal side, thereby generating a water-hammer shock. The bubble thereafter takes the form of two vortex lines convecting downstream.
The present result in Fig.~\ref{Fig:Cavity-A} with the Phase-Field mechanism agrees well with that in \citep{Shuklaetal2010} where the model of Allaire et al. \citep{Allaireetal2002} with interface sharpening was used.
When the Phase-Field mechanism is deactivated, the bubble and water are mixed as the simulation progresses, and the bubble remains connected at the end of the simulation with a ``tail'' seeming to connect the two vortex lines. However, these behaviors are not observed when including the Phase-Field mechanism.

\begin{figure}[!t]
	\centering
    \includegraphics[scale=0.3]{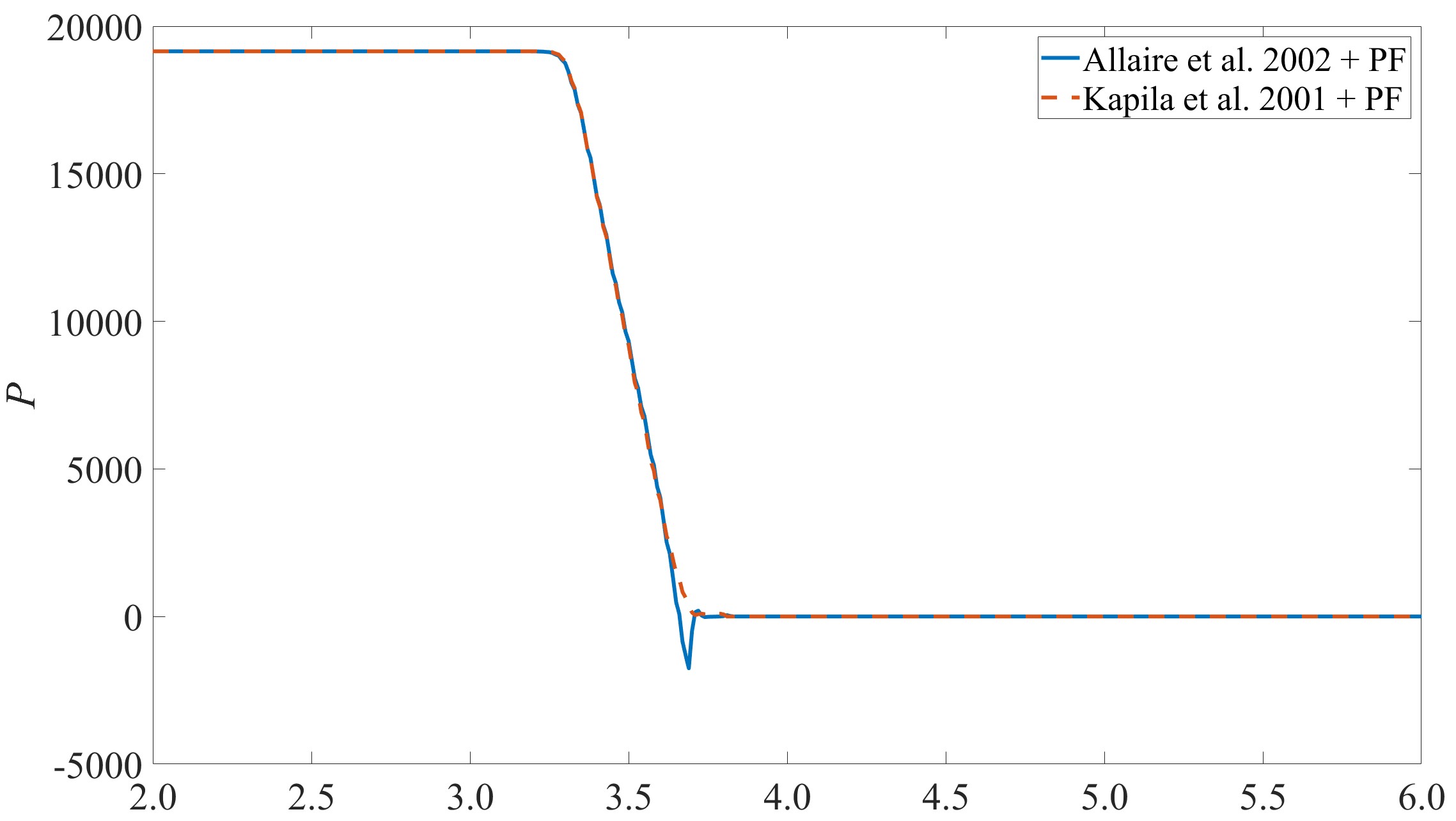}
	\caption{Pressure in the shock-induced collapse of an air cavity in water at $y=0.70$ and $t=0.010$.}\label{Fig:Cavity-Pressure}
\end{figure}

Although the models of Allaire et al. \citep{Allaireetal2002} and Kapila et al. \citep{Kapilaetal2001} produce a similar evolution of the bubble shape in Fig.~\ref{Fig:Cavity-A} and Fig.~\ref{Fig:Cavity-K}, their shock interacting with the water-air interface behaves differently.
Fig.~\ref{Fig:Cavity-Pressure} shows the pressure along a horizontal line ($y=0.70$) that crosses the upper edge of the reflected rarefaction wave after the shock impacts the bubble ($t=0.010$). The pressure from the model of Kapila et al. \citep{Kapilaetal2001} remains positive near the water-air interface, while the reflected rarefaction wave generates a strong negative pressure spike near the interface when using the model of Allaire et al. \citep{Allaireetal2002}. This negative pressure contributes to a local small sound speed that can hinder wave propagation, resulting in incorrect bubble collapse dynamics as shown in \citep{Tiwarietal2013,Schmidmayeretal2020}.

We further illustrate the significance of satisfying the admissible (bound-preserving) requirement, in particular the energy bound in Eq.~(\ref{Eq:EnergyBound}), when high-order schemes are used. Table~\ref{Table:Cavity} lists whether simulations succeed or fail to complete when the different models are solved without considering the energy bound (i.e., Eq.~(\ref{Eq:WENO-C-Energy}) is skipped during the implementation of WENO-C).
Although the mass positivity, volume fraction boundedness, and volume fraction summation to unity are satisfied, all the simulations with (Eq.~(\ref{Eq:WENO-C-Energy})-skipped) WENO-C, except that using the model of Allaire et al. \citep{Allaireetal2002} without the Phase-Field mechanism, fail within a short period of time after the shock impacts the bubble ($t\approx 0.0085$), as the energy bound is not preserved. However, using MUSCL, which is more dissipative, the simulations with all four models run to completion without the need to enforce the energy bound. This comparison highlights the crucial role played by the bound-preserving requirement for simulations of challenging compressible multiphase flows, in particular when high-order schemes are used.
\begin{table}[]
    \centering
    \begin{tabular}{c|c|cc}
    \hline
         Model&                                                  WENO5-C&                        MUSCL&\\
         \hline
         Model of Allaire et al. \citep{Allaireetal2002}&        Success&                        Success&\\
         Model of Allaire et al. \citep{Allaireetal2002} + PF&   Failure at $t=0.0096$&          Success&\\
         Model of Kapila et al. \citep{Kapilaetal2001}&          Failure at $t=0.0120$&          Success&\\
         Model of Kapila et al. \citep{Kapilaetal2001} + PF&     Failure at $t=0.0093$&          Success&\\
    \hline
    \end{tabular}
    \caption{Simulation success/failure for the shock-induced collapse of an air cavity in water without preserving the energy bound}
    \label{Table:Cavity}
\end{table}

\subsubsection{Shock–vessel–bubble interaction}\label{Sec:ShockVesselBubble}
A $40$ MPa shock in a tissue modeled by 10\% gelatin (Phase~$3$: $\gamma_3=6.72$, $P_3^{\infty}=3.70\times10^{8}$, and $D_3=0.00$) impinges on a vessel filled with water (Phase~$2$: $\gamma_2=6.12$, $P_2^{\infty}=3.43\times10^{8}$, and $D_2=0.00$); inside the vessel there is an air bubble (Phase~$1$: $\gamma_1=1.40$, $P_1^{\infty}=0.00$, and $D_1=0.00$). The inviscid case is considered without modeling the elasticity, following \citep{CoralicColonius2014}.
The initial conditions for this three-dimensional problem are
\begin{equation}\label{Eq:ShcokVesselBubble-IC}
\begin{split}
&(\rho_1,\rho_2,\rho_3,u,v,w,P,\alpha_1,\alpha_2,\alpha_3)=\\
&\left\{
\begin{array}{cc}
(1.204,1000.0,1030.0,0.0,0.0,0.0,101325.0,1,0,0), &  \psi_1 \geqslant 0,\\
(1.204,1000.0,1030.0,0.0,0.0,0.0,101325.0,0,0,1), &  \psi_3 \geqslant 0 \quad \& \quad x \leqslant x_{S},\\
(1.204,1000.0,1046.0,-24.2,0.0,0.0,4.0\times10^{7},0,0,1), &  \psi_3 \geqslant 0 \quad \& \quad x > x_{S},\\
(1.204,1000.0,1030.0,0.0,0.0,0.0,101325.0,0,1,0), & \mathrm{else},
\end{array}
\right.
\end{split}
\end{equation}
where $x_{S} = 20.0\times10^{-6}$, $\psi_1 = r_1 - \sqrt{ (x-x_1)^2 + (y-y_1)^2 + (z-z_1)^2 )}$ with $x_1=0.0$, $y_1 = 0.0$, $z_1 = 0.0$ and $r_1=10.0\times10^{-6}$, and $\psi_3 = \sqrt{ (x-x_3)^2 + (y-y_3)^2 } - r_3$ with $x_3 = 0.0$, $y_3 = 0.0$, and $r_3  = 13.0\times10^{-6}$, as shown in Fig.~\ref{Fig:Application-IC} (right).
The domain is $[-50.0\times10^{-6},50.0\times10^{-6}]\times[0.0,50.0\times10^{-6}]\times[0.0,50.0\times10^{-6}]$, covering a quarter of the problem due to symmetry, and is discretized with $400\times200\times200$ grid cells. As a result, outflow boundary conditions are used except on the symmetric planes at $y=0.0$ and $z=0.0$. All the quantities here are in their SI units.

\begin{figure}[!t]
	\centering
	\includegraphics[scale=0.33]{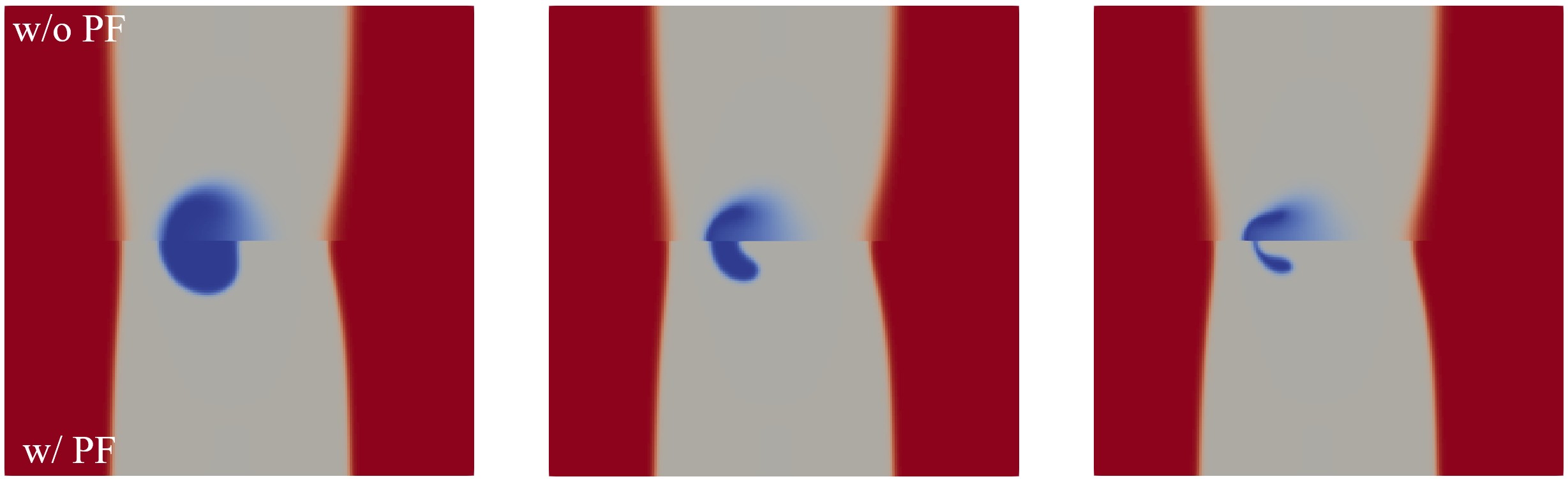}\\
    \includegraphics[scale=0.33]{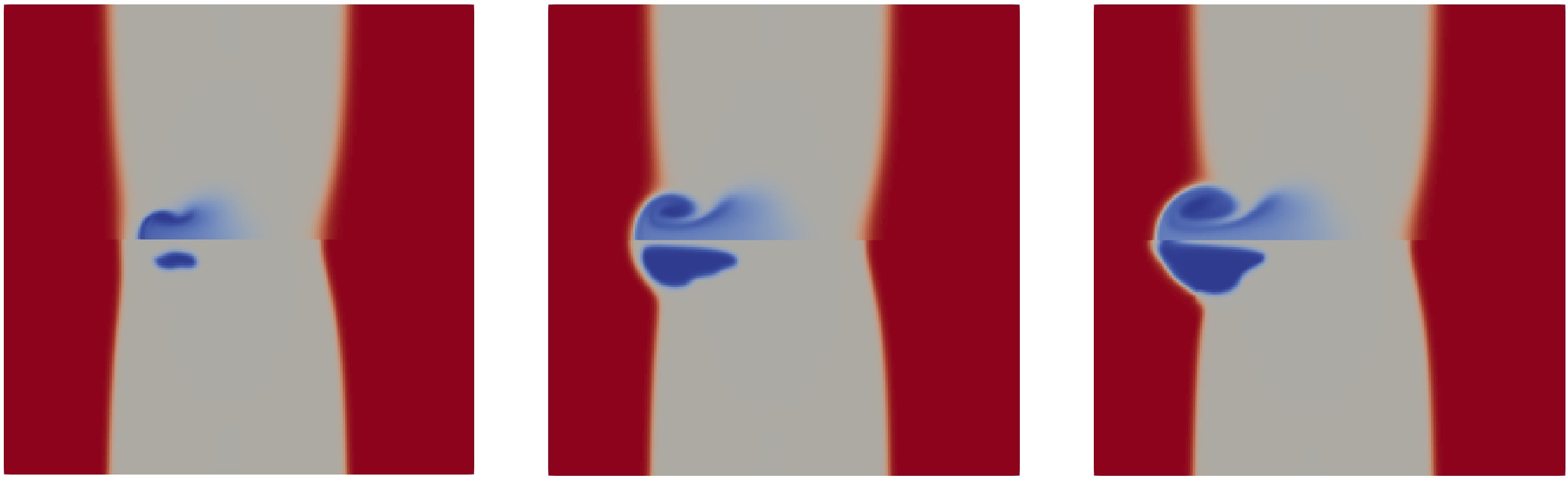}
	\caption{Evolution of the three phases ($\sum_{p=1}^N p\times \alpha_p$) in the shock-vessel-bubble interaction problem at $t=50$, $55$, $58$, $62$, $75$, and $84$ $\mathrm{ns}$ (from left to right and top to bottom) without (top frame) and with (bottom frame) the Phase-Field mechanism using the model of Allaire et al. \citep{Allaireetal2002}.}\label{Fig:ShockVesselBubble-A}
\end{figure}

\begin{figure}[!t]
	\centering
	\includegraphics[scale=0.34]{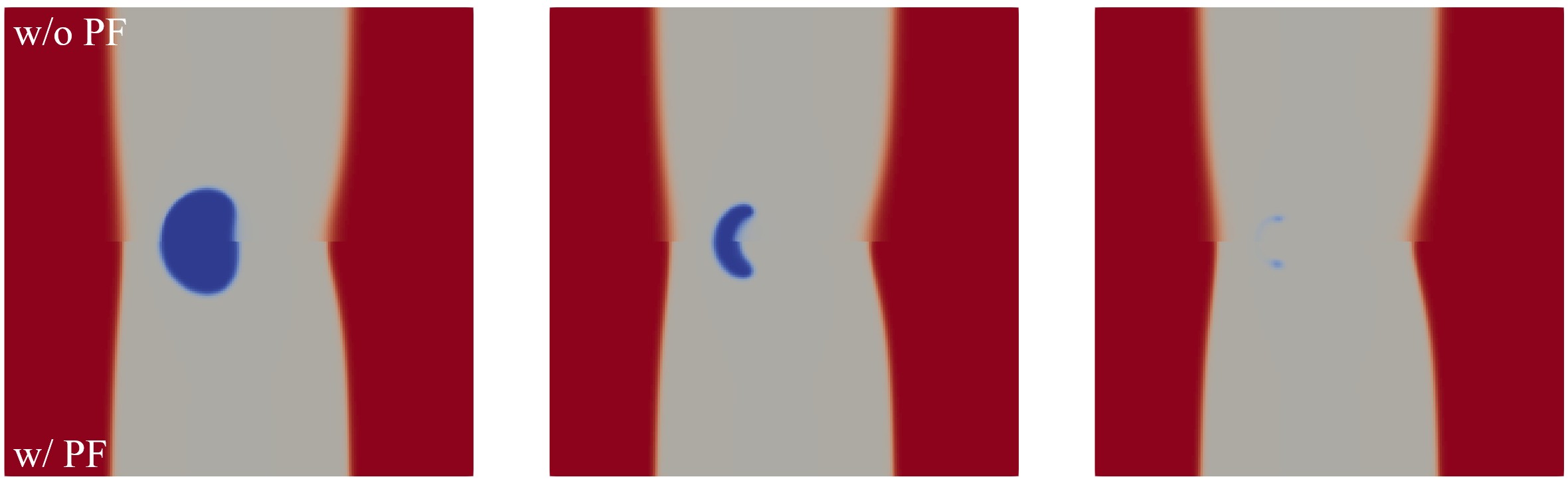}\\
    \includegraphics[scale=0.34]{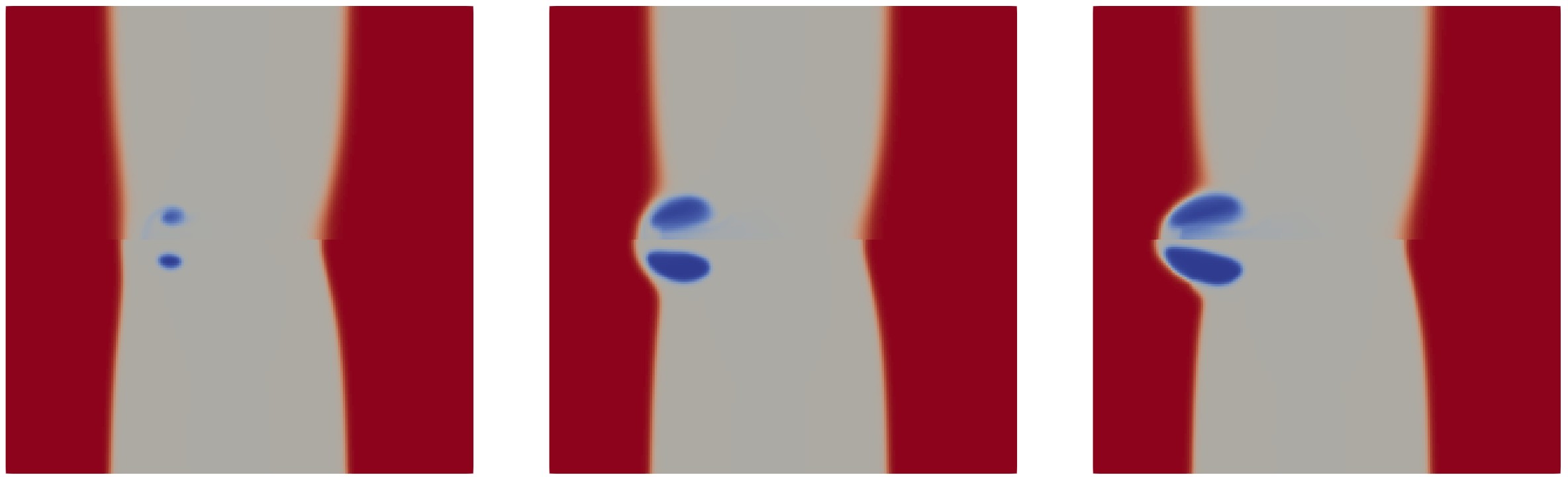}
	\caption{Evolution of the three phases ($\sum_{p=1}^N p\times \alpha_p$) in the shock-vessel-bubble interaction problem at $t=50$, $55$, $58$, $62$, $75$, and $84$ $\mathrm{ns}$ (from left to right and top to bottom) without (top frame) and with (bottom frame) the Phase-Field mechanismusing the model of Kapila et al. \citep{Kapilaetal2001}.}\label{Fig:ShockVesselBubble-K}
\end{figure}

Fig.~\ref{Fig:ShockVesselBubble-A} and Fig.~\ref{Fig:ShockVesselBubble-K} show the evolution of the three phases quantified by $\sum_{p=1}^N p\times \alpha_p$ at selected moments without and with the Phase-Field mechanism, using the models of Allaire et al. \citep{Allaireetal2002} and Kapila et al. \citep{Kapilaetal2001}, respectively.
The bubble collapses due to the impact of the transmitted shock from the gelatin to the water. A liquid jet forms towards the left edge of the vessel, breaking the bubble and convecting it toward the vessel wall. During the bubble collapse and expansion, the vessel wall is significantly deformed. The present result in Fig.~\ref{Fig:ShockVesselBubble-A} without the Phase-Field mechanism agrees with that in \citep{CoralicColonius2014} where the model of Allaire et al. \citep{Allaireetal2002} was used.
When the Phase-Field mechanism is inactive, the bubble and the water inside the vessel are mixed by numerical diffusion; it is difficult to distinguish the bubble interface.
This behavior is most evident in the results with the model of Allaire et al. \citep{Allaireetal2002}. In contrast, the bubble interface when using the model of Kapila et al. \citep{Kapilaetal2001} is much sharper even when the Phase-Field mechanism is not activated. However, there is still some amount of air numerically diffused into the water, resulting in a smaller bubble size than that with the Phase-Field mechanism after the bubble rebounds. 
Overall, including the Phase-Field mechanism successfully prevents numerical mixing of different phases; each phase can be specified unambiguously.

Again, different bubble collapse dynamics are predicted by the models of Allaire et al. \citep{Allaireetal2002} and Kapila et al. \citep{Kapilaetal2001}.
In Fig.~\ref{Fig:ShockVesselBubble-A}, the bubble from the model of Allaire et al. \citep{Allaireetal2002} has a stronger resistance to compression, and thus is more influenced by the convection of fluid flow; the bubble breaks and rotates due to the formation of a jet and finally has a ``mushroom'' shape due to the formation of a vortex ring. Different from that, in Fig.~\ref{Fig:ShockVesselBubble-K}, the bubble from the model of Kapila et al. \citep{Kapilaetal2001} is compressed to a level that the bubble is hardly visible before it starts to rebound.
Our observation here is also consistent with the comparison study in \citep{Tiwarietal2013,Schmidmayeretal2020}; the bubble from the model of Allaire et al. \citep{Allaireetal2002} was more difficult to be compressed than that from the model of Kapila et al. \citep{Kapilaetal2001}, and thus the bubble rebounded at an incorrectly large radius that disagrees with the Keller-Miksis model \citep{KellerMiksis1980} for spherical bubble collapse.

\begin{figure}[!t]
	\centering
	\includegraphics[scale=0.4]{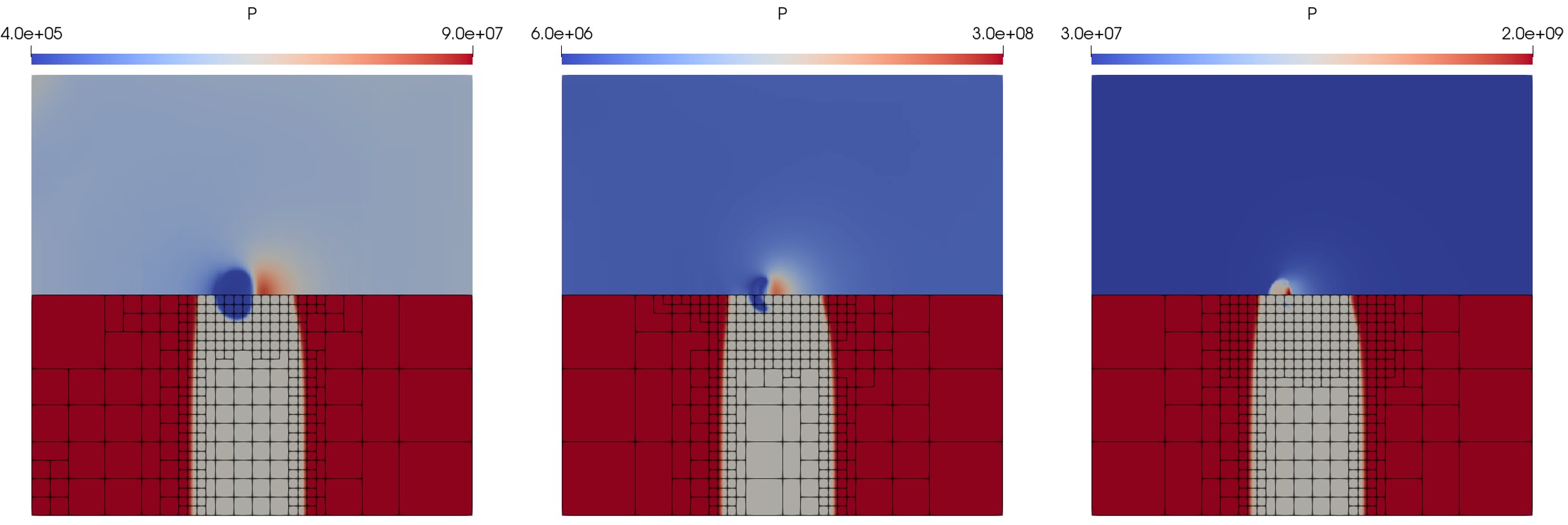}\\
        \includegraphics[scale=0.4]{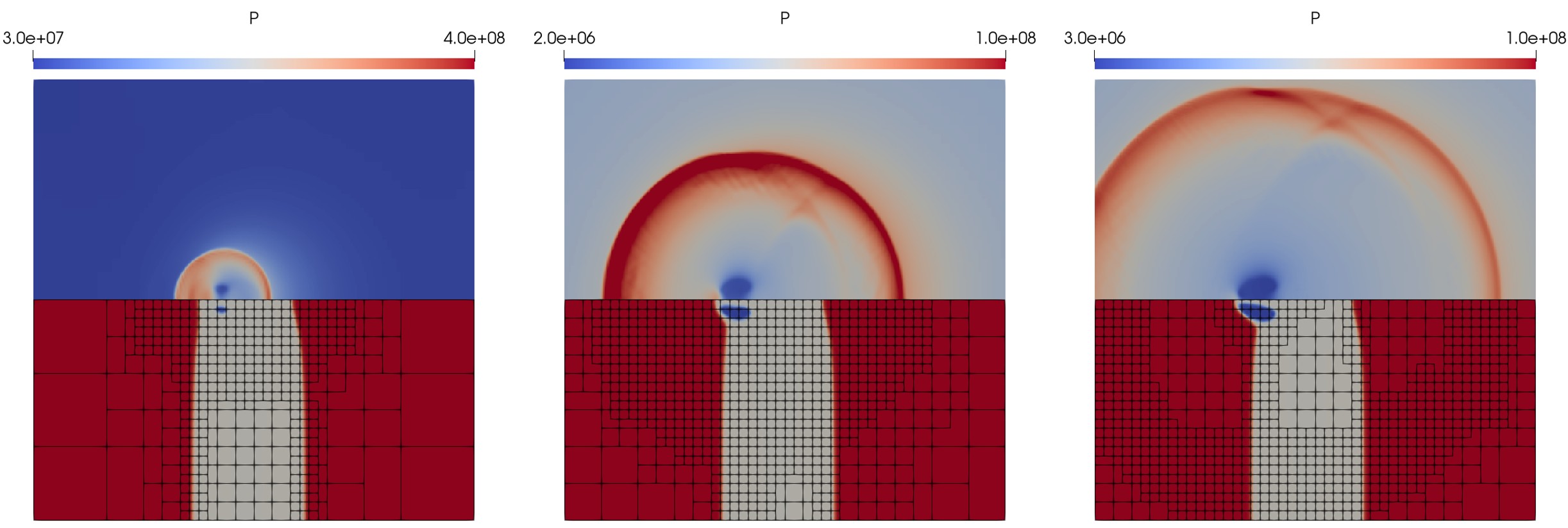}
	\caption{Evolution of the pressure (top frame) and the three phases ($\sum_{p=1}^N p\times \alpha_p$) (bottom frame) in the shock-vessel-bubble interaction problem at $t=50$, $55$, $58$, $62$, $75$, and $84$ $\mathrm{ns}$ with the model of Kapila et al. \citep{Kapilaetal2001}, the Phase-Field mechanism, and adaptive mesh refinement.}\label{Fig:ShockVesselBubble-AMR}
\end{figure}

We finally demonstrate the incorporation of the proposed WENO-C scheme with adaptive mesh refinement using a $48\times24\times24$ base mesh and 3 levels of refinement. The finest mesh size is $384\times192\times192$, close to the resolution of the $400\times200\times200$ uniform mesh.
The mesh is refined when the difference of $\sum_{p=1}^N p \times \alpha_p$ from its neighboring values is greater than $0.2$ or the difference of the pressure from its neighboring values is greater than $1.0\times10^{6}$.
Fig.~\ref{Fig:ShockVesselBubble-AMR} shows the pressure and the phases ($\sum_{p=1}^N p \times \alpha_p$) along with the outline of the AMR mesh at selected moments from the model of Kapila et al. \citep{Kapilaetal2001} with the Phase-Field mechanism. Fine meshes are dynamically allocated following the evolution of the phases and pressure. The AMR result in Fig.~\ref{Fig:ShockVesselBubble-AMR} agrees well with that from the uniform mesh in Fig.~\ref{Fig:ShockVesselBubble-K}, while the wall time is about six times less.

\section{Conclusion}\label{Sec:Conclusions}
In the present study, we propose a consistent and bound-preserving WENO reconstruction scheme (``WENO-C'') for compressible two-/$N$-phase flows with the Phase-Field mechanism.
Our proposed WENO-C scheme satisfies four inter-dependent requirements: consistency \citep{HuangJohnsen2022,HuangJohnsen2023,HuangJohnsen2024}, equilibrium \citep{Abgrall1996,JohnsenColonius2006,BeigJohnsen2015}, conservation \citep{HenrydeFrahanetal2015,HuangJohnsen2024}, and admissibility (bound preservation) \citep{HuangJohnsen2024}, which are critical for accurate compressible multiphase flow simulations. When there are more than two phases, the WENO-C scheme additionally satisfies consistency of reduction and volume fraction summation to unity, eliminating any numerical production of fictitious phases, local voids, or overfilling.
The WENO-C scheme relies on a modified calculation of WENO weights that are based on the relative smoothness between stencil candidates, and a coupled reconstruction strategy for the masses and volume fractions that specify locations of material interfaces. These modifications have no effect on the order of accuracy in smooth problems.
Our analysis illustrates that the standard WENO-JS scheme \citep{JiangShu1996} fails to maintain thermal equilibrium (when it should) and generates spurious oscillations in bulk phases as well as at interfaces of low‑density phases. These errors are reduced to round-off, and both shocks and material interfaces are well captured by the proposed WENO-C scheme.
All the properties mentioned are carefully verified with numerical experiments, after implementing the proposed WENO-C scheme with the consistent and conservative Phase-Field method for compressible multiphase flows that enables adaptive mesh refinement \citep{HuangJohnsen2022,HuangJohnsen2023,HuangJohnsen2024,Huangetal2025}.

The proposed WENO-C scheme is applied to simulate the shock-induced collapse of an air cavity in water \citep{Shuklaetal2010} (two-dimensional, two-phase) and shock-vessel-bubble interaction \citep{CoralicColonius2014} (three-dimensional, three-phase), which were previously investigated with the model of Allaire et al. \citep{Allaireetal2002}. 
However, comparison studies \citep{Tiwarietal2013,Schmidmayeretal2020} show that the model of Kapila et al. \citep{Kapilaetal2001}, not Allaire et al. \citep{Allaireetal2002}, produces correct bubble collapse dynamics. In the present study, both the models of Allaire et al. \citep{Allaireetal2002} and Kapila et al. \citep{Kapilaetal2001} without and with the Phase-Field mechanism are successfully solved with the proposed WENO-C scheme, and, as a result, different behaviors of the two models are further illustrated.
In comparison to the model of Kapila et al. \citep{Kapilaetal2001}, the model of Allaire et al. \citep{Allaireetal2002} has a more significant numerical mixing among different phases and allows a more negative pressure to appear near water-air interfaces, resulting in low sound speed regions that can affect wave propagation and the rate of bubble collapse. Our result is consistent with previous observations \citep{Shuklaetal2010,CoralicColonius2014,Tiwarietal2013,Schmidmayeretal2020}.
For both models, the Phase-Field mechanism effectively prevents mixing of different phases from numerical diffusion.

During the numerical investigation, we also compare the convergence rate with respect to mesh size in a problem with shocks and interfaces, and the theoretically optimal convergence rate \citep{LeVeque2002} is achieved after including the Phase-Field mechanism due to its ability to maintain a constant interface thickness in the order of the mesh size.
We observe simulation failures after shocks impact interfaces when the energy bound \citep{HuangJohnsen2024} of WENO-C is deactivated, demonstrating the important role of the bound-preserving property for high-order schemes in compressible multiphase flow simulations.
The incorporation of WENO-C with adaptive mesh refinement is straightforward, and a significant reduction in simulation time is reported without sacrificing accuracy when AMR is enabled.
We further discuss WENO-CL, a variant of the proposed WENO-C scheme using the consistent limiter \citep{HuangJohnsen2023}. An additional property of WENO-CL is maintaining a non-zero and non-unity constant volume fraction, a scenario that is impossible for immiscible phases but can occur in problems like combustion. Although applied to volume fractions, WENO-CL suggests clear potential for miscible multiphase problems where mass fractions are used more frequently; this will be examined in future research.

\section*{Acknowledgments}
ZH acknowledges Prof. Eric Johnsen and Dr. William J. White for their fruitful discussions while ZH was at the University of Michigan.

\appendix
\section{Accuracy of WENO-CL}\label{Appendix:WENO-CL-Accuracy}
\begin{figure}[!t]
	\centering
    \includegraphics[scale=0.33]{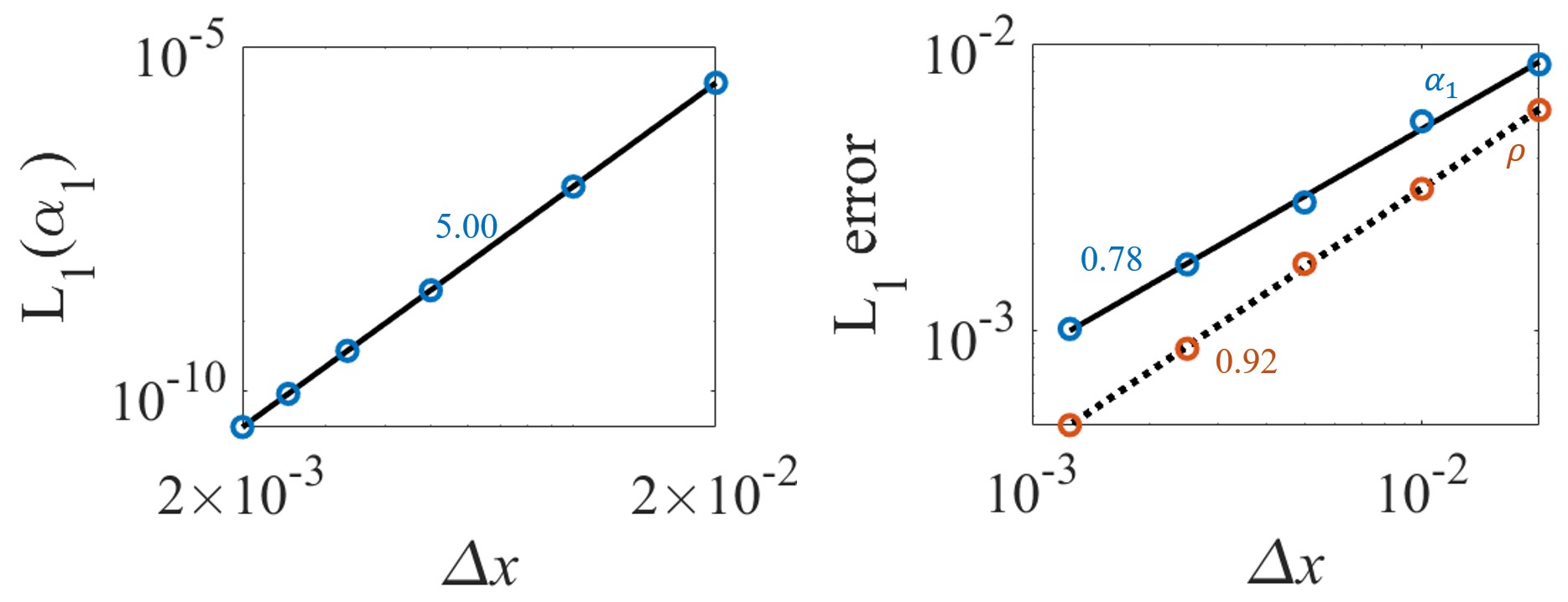}
	\caption{Volume fraction error in the smooth problem in Section~\ref{Sec:Advection-Smooth} (left), and volume fraction and density errors in the discontinuous problem in Section~\ref{Sec:AirHeliumShock} (right), with WENO5-CL.}\label{Fig:WENO-CL-Accuracy}
\end{figure}

The problems in Section~\ref{Sec:Advection-Smooth} and Section~\ref{Sec:AirHeliumShock} are repeated using WENO-CL in Section~\ref{Section:Limiter-Consistent} to verify its accuracy.
Fig.~\ref{Fig:WENO-CL-Accuracy} shows the errors with respect to the mesh size, along with the fitted convergence rate. A fifth-order convergence rate is observed in the smooth problem (Section~\ref{Sec:Advection-Smooth}), while convergence rates similar to those in Fig.~\ref{Fig:AirHelium-L1} are obtained in the discontinuous problem (Section~\ref{Sec:AirHeliumShock}). The result demonstrates that the present implementation of the consistent limiter \citep{HuangJohnsen2023} to high-order schemes does not affect the order of accuracy in both smooth and discontinuous problems.

\bibliographystyle{plain}
\bibliography{refs.bib}

\end{document}